\documentclass[11pt,draftclsnofoot,onecolumn,journal]{IEEEtran}
\usepackage{amsmath,amsfonts}
\usepackage{algorithmic}
\usepackage{algorithm}
\usepackage{array}
\usepackage{textcomp}
\usepackage{stfloats}
\usepackage{url}
\usepackage{verbatim}
\usepackage{graphicx}
\usepackage{cite}
\usepackage{hyperref}
\usepackage{booktabs}
\usepackage{environ}
\usepackage{wrapfig}
\usepackage{subcaption}

\usepackage{xcolor}
\usepackage[normalem]{ulem}

\NewEnviron{FitToWidth}[1][\textwidth]{%
\begin{center}
  \makebox[#1]{\resizebox{#1}{!}{\BODY}}%
\end{center}
}
\begin{document}

\title{Automated Analysis of Naturalistic Recordings in Early Childhood: Applications, Challenges, and Opportunities}

\author{Jialu Li~\IEEEmembership{Member,~IEEE}, Marvin Lavechin, Xulin Fan~\IEEEmembership{Student Member,~IEEE},  Nancy L. McElwain, Alejandrina Cristia, Paola Garcia-Perera~\IEEEmembership{Member,~IEEE}, Mark Hasegawa-Johnson~\IEEEmembership{Fellow,~IEEE}} 

\markboth{IEEE Signal Processing Magazine}%
{\MakeLowercase{\textit{et al.}}: Author Guidelines for Special Issue Articles of IEEE SPM}

\maketitle


Naturalistic recordings capture audio in real-world environments where participants behave naturally without interference from researchers or experimental protocols. Naturalistic long-form recordings extend this concept by capturing spontaneous and continuous interactions over extended periods, often spanning hours or even days, in participants' daily lives.
Naturalistic recordings have been extensively used to study children's behaviors, including how they interact with others in their environment, in the fields of psychology, education, cognitive science, and clinical research. These recordings provide an unobtrusive way to observe children in real-world settings beyond controlled and constrained experimental environments.
Advancements in speech technology and machine learning have provided an initial step for researchers to automatically and systematically analyze large-scale naturalistic recordings of children. Despite the imperfect accuracy of machine learning models, these tools still offer valuable opportunities to uncover important insights into children's cognitive and social development.
Several critical speech technologies involved include speaker diarization, vocalization classification, word count estimate from adults, speaker verification, and language diarization for code-switching. Most of these technologies have been primarily developed for adults, and speech technologies applied to children specifically are still vastly under-explored. To fill this gap, we discuss current progress, challenges, and opportunities in advancing these technologies to analyze naturalistic recordings of children during early development ($<3$ years of age). We strive to inspire the signal processing community and foster interdisciplinary collaborations to further develop this emerging technology and address its unique challenges and opportunities.


\section*{Introduction}
\label{sec:scope}

Naturalistic recordings offer valuable insights into various aspects of child development, including understanding adult-child interactions, monitoring language acquisition, and early detection of disorders and developmental delays. As digital audio recorder technology has evolved, it has become easier to collect audio data from children in their daily lives. This has led to significant advancements in research on language acquisition and social behavior. Along with the advancement of speech processing technology on automatic analysis of naturalistic recordings, researchers can systematically analyze large-scale daylong naturalistic recordings of children across different countries, cultures, and languages~\cite{bergelson2023everyday}. 

In this article, we will review the current progress, challenges, and opportunities in several key speech processing tasks for automatically analyzing naturalistic audio collected from infancy to early childhood (aged 0-3 years). 
These tasks have been mainly tested on child-centered naturalistic long-form recordings in previous studies, including \textit{speaker diarization} or \textit{speaker type classification} to identify who spoke when, \textit{vocalization classification} to monitor the emotional states of both adults and children, children's \textit{vocal maturity classification} to assess developmental progress, and \textit{adult word/phoneme count estimation} to quantify children's language exposure. 
In line with understanding child development, we will also review two emerging tasks recently explored: \textit{speaker verification of newborn crying} collected in hospitals and \textit{language diarization and identification} within code-switching families at home, both of which have the potential for future application to child-centered naturalistic long-form recordings. 
Besides speech processing tasks, understanding the child's acoustic environment is another important aspect of audio processing. Although environmental sound classification in domestic settings can provide insights into a child's surroundings, this task falls within the broader field of acoustic event detection, which has been thoroughly reviewed in the literature. Thus, this article will focus on speech-related processing rather than general environmental sound analysis.
Many of the above speech processing tasks have been primarily developed using data from adults, and speech processing technology developed for children remains significantly under-explored. Previous review papers related to children often focus on one specific task, such as automatic speech recognition (ASR) for children over five years of age~\cite{bhardwaj2022automatic}, automatic detection of speech disorders in children over three years of age~\cite{8931568}, and infant crying classification and analysis~\cite{ji2021review}. 
This paper aims to contribute to this literature by presenting a comprehensive survey of robust speech processing algorithms developed for analyzing naturalistic recordings of children under three years of age from a signal processing perspective.

These speech-processing algorithms hold great potential for two distinct but related reasons. On the one hand, describing children's everyday language input and outcomes can help us build scientific knowledge, contributing to disparate fields like anthropology and cognitive science. Most children go from mere coos to complex verbal coordination in just a few years, and understanding the stages of this development and the cognitive mechanisms subtending it is a key scientific enterprise. Algorithmic approaches applied to long-form recordings of infants in their natural environments can play a crucial role in contributing to this body of knowledge. On the other hand, these algorithms are essential to building artificial intelligence (AI) applications to support better social, emotional, and cognitive outcomes for children. For example, automatic analysis of naturalistic speech may help in the early detection of critical events associated with developmental or mental disorders in children. These events include low frequency of child vocalizations, excessive crying, and frequent uncoordinated parent-child vocal interactions. 

\section*{Data Collection Procedures and Considerations}
\label{sec:length}
Although the focus of this review is on advances in speech technology and machine learning (ML) algorithms, we first provide a brief overview of standard procedures and considerations when collecting long-form audio recordings among infants and young children in home environments (see~\cite{levin2021sensing} and~\cite{mcelwain2024evaluating} for more detailed discussions).
\subsection*{Recording Devices} 
\begin{figure}[h]
	\centering
	\includegraphics[width=1.0\textwidth]{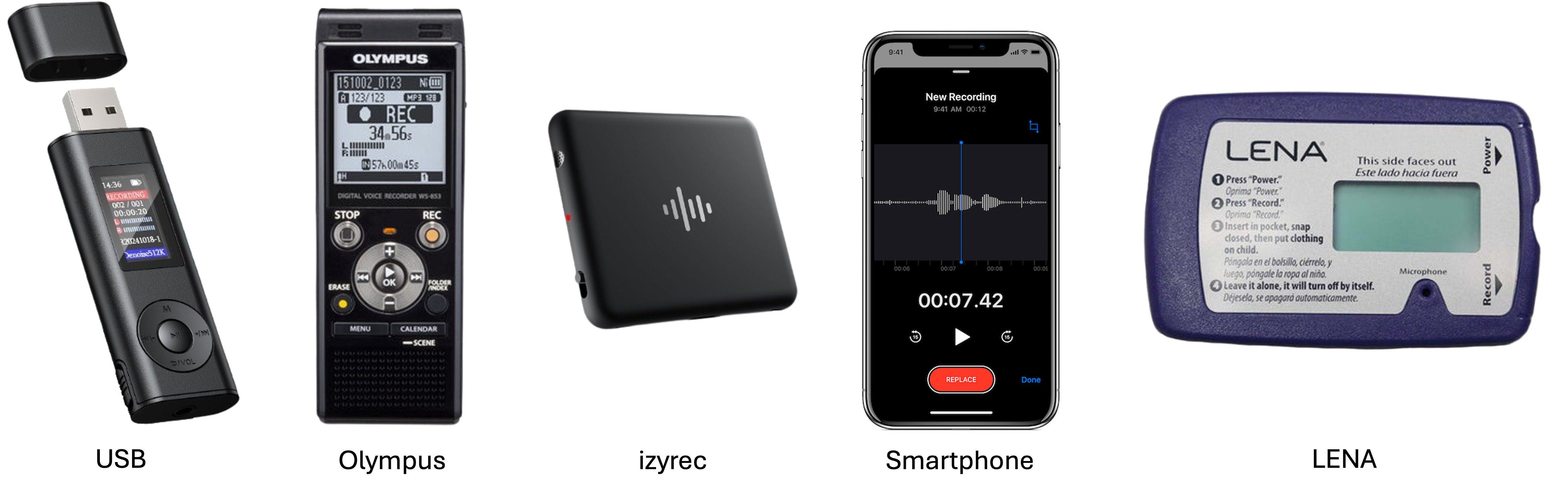}
	\caption{{Common devices for everyday egocentric audio recording.}}
	\label{fig:recording_device}
\end{figure}

In typical data collection pipelines, the child wears an audio recorder throughout the day to capture an egocentric view of their daily experiences. The devices used to record naturalistic audio range from all-purpose audio recorders, such as USB, Olympus, and izyrec, to specifically designed child-wearable devices equipped with proprietary speech technology, such as Language ENvironment Analysis (LENA)~\cite{gilkerson2008lena}.
Although all-purpose audio recorders provide budget-friendly and flexible solutions for recording long-form audio, they lack the functionality of automatic analysis of child vocalizations and child-adult vocal interactions. In addition to these dedicated wearable devices, smartphones are increasingly a practical alternative for naturalistic recording. Modern smartphones are equipped with high-quality microphones, support a variety of recording apps, and their ubiquity makes them especially appealing for studies conducted in resource-constrained environments. 
However, smartphones may be less effective for long-form recordings, as continuous audio recording can interfere with regular device use or be interrupted by other app usage. Moreover, when used solely for audio recording, smartphones are a relatively high-cost option compared to other recording devices.
Figure~\ref{fig:recording_device} shows different types of wearable recording devices.
Table~\ref{table:device} provides a high-level overview of approximate prices and key features of different wearable recorders.

LENA is a commercially available device that can be worn by the child in a pocket on a specially designed shirt, recording up to 24 hours of audio.
LENA provides an end-to-end speech processing pipeline that identifies speakers, recognizes environmental sounds, and measures both the language input children receive and the sounds they produce. These measures include the number of vocal turns between an adult speaker and the key child, the number of words spoken by adult speakers, and the number of child vocalizations. 
LENA's ease of use, along with its proprietary speech processing technology for automated analysis of children's language environments, has made it popular within the research community. One systematic review of 53 studies highlighted several applications of the LENA system, including assessment of individual differences in child language exposure and production, as well as mean differences in language exposure or production as a function of language spoken in the home (e.g., English versus Spanish), the child’s developmental status (e.g., typically development versus developmentally delayed), and the recording context (e.g., home versus classroom)~\cite{greenwood2018automated}.
However, LENA's closed-source algorithms prevent researchers from accessing its implementations or applying them to external audio files collected via other recorders. 


\begin{table}[H]
	\centering
	\setlength{\tabcolsep}{2.0pt}
	\begin{FitToWidth}[\columnwidth]
		\begin{tabular}{|l|l|l|l}
			\hline
			\textbf{Recorder} & \textbf{Price} & \textbf{Key Features} \\ \hline
			USB recorder & \$5-\$100 & Portable, direct USB connection, easy data transfer\\
            \hline
			Olympus & \$140 &  High-quality professional recordings, relatively heavy. \\
            \hline
			izyrec & \$100 & App integration, AI noise canceling, long duration (30-hour) recording time. \\
            \hline
            {Smartphone} & {\$300-\$1200} & {Ubiquitous and multi-functional; supports high-quality recording but may be} \\
            && {limited for long-form use.}\\
            \hline
			LENA & \$219-\$329 & 24-hour recording, proprietary technology with language environment analysis, \\
            & {(device only)} & {which requires an additional paid subscription.} \\
		\hline
		\end{tabular}
	\end{FitToWidth}
	\vspace{0.1cm}
	\caption{Comparisons among audio recording devices, including approximate prices and key features.}
	\label{table:device}
	\vspace{-0.5cm}
\end{table}

\subsection*{Data Annotations}

There are two primary types of data annotation for naturalistic recordings: \textit{manual coding} performed by trained experts and \textit{crowd-sourcing} by individuals without prior training who perform simple labeling tasks.

\begin{figure}
    \centering
    \includegraphics[width=1.0\linewidth]{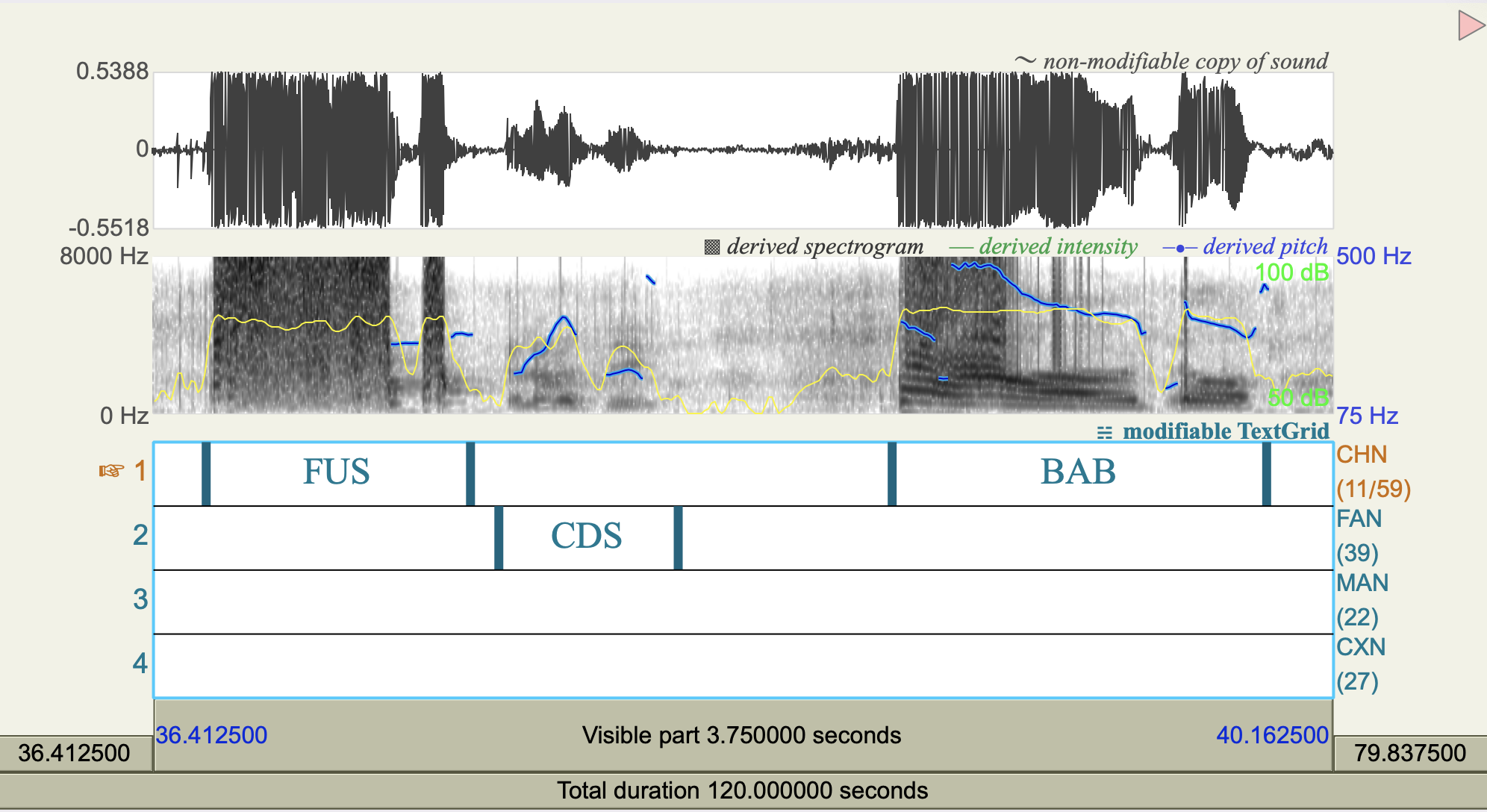}
	\caption{A Praat window of sample labeled segments of four speakers: the key child who wears the recorder (CHN), adult female (FAN), adult male (MAN), and other child or sibling (CXN). The CHN fusses (FUS) and babbles (BAB), while FAN speaks in child-directed speech (CDS).}
	\label{fig:praat_window}
\end{figure}

\textit{Manual coding} by trained experts is considered as ground truth and represents the state-of-the-art human performance in understanding the full acoustic scenes captured in naturalistic recordings. However, data annotation is labor- and time-intensive. It is impractical for researchers to annotate all the collected long-form recordings. Typically, trained annotators code snippets of relevant audio segments, e.g., sections with high levels of speech activity according to an algorithm. Small numbers of files are typically double-annotated, and inter-rater reliability scores are calculated to ensure high quality of coded data. Trained coders often use professional audio annotation software, such as Praat, ELAN (EUDICO Linguistic Annotator), and Audacity, to label the selected segments.
Figure~\ref{fig:praat_window} shows a sample Praat window of labeled speaker and vocalization types in a naturalistic home recording. 

Despite its precision, manual coding significantly limits the scale and duration of audio analysis. This constraint highlights the critical need for developing robust signal processing algorithms capable of systematically analyzing large-scale naturalistic recordings. Such algorithms have the potential to uncover complex child-adult interaction patterns and yield innovative findings.

\textit{Crowd-sourcing} platforms, in contrast, are a relatively inexpensive way to obtain large-scale labeled datasets. Individual contributors may use their common sense knowledge to perform simple annotation tasks. On Zooniverse, a citizen science platform, Semenzin et al.~\cite{semenzin2021describing} asked volunteers to classify a large-scale dataset of 20 toddler vocalizations as one of five categories: canonical (contains consonant-vowel transitions), non-canonical (does not contain consonant-vowel transitions), crying, laughing, and junk. 
The annotations for each audio clip were determined by the majority vote of five volunteer coders and then compared to expert annotations. The classifications from the Zooniverse platform achieved a moderate accuracy of 67\% and an unweighted average recall (UAR) of 58.9\%. More importantly, the study found a strong correlation between expert and crowd-sourced annotations per child in two key measures: linguistic proportion, the ratio of linguistic vocalizations (canonical + non-canonical) to all vocalizations, and canonical proportion, the ratio of vocalizations containing canonical transitions or syllables to all linguistic vocalizations. These findings show that crowd-sourcing has the potential to help researchers obtain large-scale datasets of children's vocalizations at a lower cost.

\subsection*{Ethics} 

Collection of long-form audio recordings among young children in home environments requires careful consideration of ethical issues. Two key issues are participant privacy, defined as a participant’s right to control access to their personal information, and data confidentiality, defined as a researcher’s obligation to protect the participant's data in accordance with the participant's wishes. Informed consent serves as the foundation for establishing and protecting participant privacy. As part of the consent process, parents/guardians are provided with information about the study’s procedures, risks, and benefits in accessible language (typically 6th-8th grade reading level) and have the opportunity to ask questions and receive additional information needed to make an informed decision about participation. With respect to privacy concerns surrounding home audio recordings specifically, parents should be given control over these data to the greatest extent possible. For example, parents are informed that they are free to turn off or pause the recording device at any time and/or that they can request that their recordings be partially or fully destroyed and not used in the research. Parents/guardians should also be told how their data will be protected by the researcher, including who will have access, how data will be stored, and whether or how data will be shared. How researchers protect against loss of data confidentiality will depend on procedures related to how data are transferred from the recording device to the laboratory, how annotations of the raw audio data are conducted and by whom, and how and where data are stored. Because children receive special protections as research participants, a notable exception to privacy/data confidentiality is made with regard to mandated reporting. That is, parents are informed that if there is disclosure or indication of actual or suspected abuse, neglect, or exploitation of a child, the researchers are required to report the information to the relevant authorities. Prior reports have addressed these and other ethical issues related to long-form home audio recordings from both researchers'~\cite{cychosz2020longform} and participants'~\cite{levin2021sensing, mcelwain2024evaluating} perspectives. In sum, ethical issues must be carefully considered and addressed in any research program that involves collecting and/or analyzing data from long-form audio recordings collected from children and parents in home environments.

\subsection*{Data Repositories}

Due to ethical and privacy concerns, data-sharing of naturalistic recordings with the public is limited.
To address the challenge, TalkBank~\cite{macwhinney2007talkbank} aims to foster research on human communication with an emphasis on spoken communication through open data-sharing. 
Within TalkBank, key sub-banks include CHILDES, which contains data on child language development in 42 languages from infancy to age 6; HomeBank, which focuses on sharing naturalistic long-form home recordings of children’s everyday experiences; and PhonBank, which analyzes children’s early phonological development in 18 languages. These sub-banks contain critical datasets for advancing speech technologies to automatically analyze naturalistic recordings.
Besides TalkBank, Databrary~\cite{simon2015databrary} provides a restricted-access data library specialized in data storage, streaming, and sharing of video and audio recordings collected as research data. Both platforms often require researchers to sign data agreements to access the data.
Both repositories host hundreds of small speech or video corpora of children collected for various study purposes, yet the variability of annotation protocols across different corpora may pose challenges for developing speech technologies that require large-scale datasets with consistent transcripts or annotations.

Open-sourced audio datasets, such as Google's AudioSet or public repositories with audio captions like FreeSound (\url{https://freesound.org/}) and BigSoundBank(\url{https://bigsoundbank.com/}), often provide relevant labeled audio events recorded with different devices under various contexts (e.g., baby cries and laughter). However, these datasets are typically \textit{weakly labeled}, often lacking precise time stamps or consistent annotation protocols. Despite these limitations, researchers may find it useful to incorporate samples of these datasets for data augmentation, which can increase the diversity of training data and thus enhance model performance. The recent release of the Ego4D dataset, a large-scale collection of audio-visual egocentric recordings of adults' daily activities, provides comprehensive benchmarks for multiple tasks including speaker diarization and social interaction analysis. This dataset may inspire future innovations in adapting algorithms designed for analyzing adult-egocentric recordings to child-egocentric recordings.

\section*{Challenges of Analyzing the Naturalistic Long-Form Recordings}
\subsection*{Noisy and Overlapped Speech} 
Naturalistic recordings often contain noisy and distorted audio due to unpredictable background sounds like household appliances, TV, street noise, and other ambient sounds that interfere with the detection and labeling of adult and child vocalizations. Additionally, far-field speech from nearby adults or older siblings can result in inconsistent speech signal quality, making it difficult for signal processing algorithms to differentiate between background noise and speech. Lastly, spontaneous adult-child interactions may involve frequent interruptions by children, leading to a noticeable amount of overlapped speech and short speaker turns. For example, the naturalistic family long-form recordings of 6- to 18-month-old infants in the SEEDLingS corpus~\cite{seedlings}, as detailed in Table~\ref{table:challenges}, contain 7.5\% overlapping speech. 
Overlapped speech and short speaker turns pose further challenges for speaker diarization.

\subsection*{Data Sparsity, Ambiguity, and Imbalanced Distributions}
Training robust ML models requires a large amount of accurately labeled datasets, but challenges arise from data sparsity, ambiguity, and imbalanced distributions. 
Around 70\% of the audio in long-form naturalistic recordings is not speech~\cite{cristia2021thorough}, and only a minority part of long-form recordings contain speech events of interest. 
Labeling audio data is an extremely time- and labor-intensive task, so obtaining accurate large-scale labeled datasets of children's audio is difficult.
Additionally, naturalistic home audio is corrupted by various types of noise, overlapping speech, and far-field reverberation and echo, and even trained coders may struggle to fully comprehend and annotate certain label types. The nuanced ambiguity of vocalization types further contributes to uncertainty when coders annotate human vocalizations under different acoustic contexts, such as distinguishing between infant crying versus fussing or between a caregiver's child-directed speech (CDS) versus adult-directed speech (ADS). Further, the imbalanced heavy-tail distribution of labels makes training ML models difficult. For example, in family audio, most vocalizations come from the target infant and adult female caregiver. The majority of infant vocalizations involve babbling, and other speakers and vocalization types make up only a small portion of the labeled data.

\subsection*{Lack of Child-Centered Naturalistic Recording Dataset for Speech Enhancement}
Mital et al.~\cite{mital2019speech} have previously explored using core digital signal processing techniques, including spectral subtraction and Wiener filtering, and neural network-based methods for speech enhancement in child-centered naturalistic home recordings. However, these approaches provided limited benefits for downstream tasks, such as distinguishing caregiver's CDS from ADS and estimating adult word count. 
This preliminary study highlights adapting existing speech enhancement techniques to naturalistic long-form recordings is challenging. Several factors may contribute to these difficulties. First, most current models are trained on adult speech datasets recorded in controlled environments. 
Given the acoustic differences between adult and infant voices, these models are likely to misclassify a majority of infant vocalizations as noise without additional fine-tuning. Additionally, the noise types commonly used in prior speech enhancement studies may not adequately represent those encountered in naturalistic family recordings. For instance, if the recording device is child-wearable, the audio may include significant noise from clothing rubbing against the microphone, a scenario not addressed in major noise datasets. Finally, the model must handle a wide range of signal-to-noise ratios (SNRs) because caregivers naturally move closer to or farther from the microphone, causing substantial variations in the target speech signal's magnitude. 


\begin{table}[]
\setlength{\tabcolsep}{1.0pt}
\resizebox{\textwidth}{!}{%
\begin{tabular}{|l|l|l|}
\hline
\textbf{Task and Label} &
  \textbf{ML Challenge} &
  \textbf{Dataset Descriptions} \\ \hline
\begin{tabular}[c]{@{}l@{}}\textbf{Speaker Diarization}\\ \textit{(time-stamped speaker activity)}\end{tabular} &
  \begin{tabular}[c]{@{}l@{}}DIHARD Challenge \\ I \& II~\cite{ryant19_interspeech}\end{tabular} &
  \begin{tabular}[c]{@{}l@{}}\textbf{SEEDLingS}~\cite{seedlings} dataset contains 1.92 hours of 23 naturalistic\\ family recordings of infants aged 6–18 months.\end{tabular} \\ \hline
\begin{tabular}[c]{@{}l@{}}\textbf{Infant Crying Classification}\\ \textit{(neutral/positive, fussing, and crying)}\end{tabular} &
  \begin{tabular}[c]{@{}l@{}}2018 IS \textsc{ComParE} \\ Crying \\ Sub-Challenge~\cite{schuller18_interspeech}\end{tabular} &
  \begin{tabular}[c]{@{}l@{}}\textbf{CRIED}~\cite{marschik2017novel} dataset contains 5587 video-taped vocalization clips recorded\\ in a laboratory setting from 20 healthy infants aged 1-4 months.\end{tabular} \\ \hline
\begin{tabular}[c]{@{}l@{}}\textbf{Toddler Vocal Maturity Classification}\\ \textit{(canonical babble, non-canonical babble,} \\ \textit{laughing, crying, junk)}\end{tabular} &
  \begin{tabular}[c]{@{}l@{}}2019 IS \textsc{ComParE} \\ Baby-Sound \\ Sub-Challenge~\cite{schuller19_interspeech}\end{tabular} &
  \begin{tabular}[c]{@{}l@{}}\textbf{BabbleCor}~\cite{cychosz2019babblecor} dataset contains 12,445 vocalization clips from naturalistic\\ long-form home recordings of 46 infants aged 2-36 months.\end{tabular} \\ \hline
\begin{tabular}[c]{@{}l@{}}\textbf{Toddler Vocal Maturity Classification}\\ \textit{(canonical babble, non-canonical babble, }\\ \textit{laughing, crying, junk)}\end{tabular} &
  - &
  \begin{tabular}[c]{@{}l@{}}\textbf{Speech Maturity} ~\cite{hitczenkospeech} dataset, an extension of BabbleCor, contains 258,914\\ clips from 398 children aged 2 months to 6 years old across 25 languages.\end{tabular} \\ \hline
\begin{tabular}[c]{@{}l@{}}\textbf{Adult Addressee Classification}\\ \textit{(infant-directed speech, }\\ \textit{adult-directed speech)}\end{tabular}&
  \begin{tabular}[c]{@{}l@{}}2017 IS \textsc{ComParE} \\ Addressee \\ Sub-Challenge~\cite{schuller17_interspeech}\end{tabular} &
  \begin{tabular}[c]{@{}l@{}}\textbf{HB-CHAAC}~\cite{schuller17_interspeech} dataset contains 10,886 vocalization clips from naturalistic\\ long-form recordings of adults interacting with infants aged 2–19 months, \\ sourced from four sub-corpora in the HomeBank repository.\end{tabular} \\ \hline
\begin{tabular}[c]{@{}l@{}}\textbf{Speaker Diarization and ASR for Children}\\ \textit{(speech transcripts, speaker diarization,} \\ \textit{human pose estimation)}\end{tabular} &
  - &
  \begin{tabular}[c]{@{}l@{}}\textbf{BabyView}~\cite{Long2024TheBD} dataset contains 493 hours of egocentric videos collected via \\ head-mounted cameras from children aged 6 months to 5 years in home \\ and preschool environments.\end{tabular} \\ \hline
\begin{tabular}[c]{@{}l@{}}\textbf{Toddler Phoneme Recognition}\\ \textit{(transcribed phoneme)}\end{tabular} &
  - &
  \begin{tabular}[c]{@{}l@{}}\textbf{Providence}~\cite{demuth2006word} corpus contains 364 hours of recordings from 6 monolingual \\ English-speaking children aged 1-3 years during spontaneous interactions \\ with their parents at home, sourced from the PhonBank repository.\end{tabular} \\ \hline
\begin{tabular}[c]{@{}l@{}}\textbf{Speaker Verification of Infant Cry}\\ \textit{(crying clips with speaker labels)}\end{tabular} &
  CryCeleb~\cite{budaghyan2024cryceleb} &
  \begin{tabular}[c]{@{}l@{}}\textbf{CryCeleb}~\cite{budaghyan2024cryceleb} dataset contains 26k audio clips totaling 6 hours of crying  \\ sounds from 786 newborns, recorded using smartphones in hospitals.\end{tabular} \\ \hline
\begin{tabular}[c]{@{}l@{}}\textbf{Language Identification and Diarization}\\ \textit{(time-stamped language labels }\\ \textit{and transcriptions)}\end{tabular} &
  MERLIon CCS~\cite{chua23_interspeech} &
  \begin{tabular}[c]{@{}l@{}}\textbf{MERLIon CCS}~\cite{chua23_interspeech} dataset contains 30 hours of recordings from over 300 \\ Zoom videos of parent-child shared book reading sessions, involving children  \\ under the age of 5 in Mandarin-English code-switching environments.\end{tabular} \\ \hline
\end{tabular}%
}
\caption{
Overview of important datasets, their associated tasks, corresponding labels, and any relevant ML challenges (if any). IS \textsc{ComParE}= Interspeech Computational Paralinguistic Challenge. } 
\label{table:challenges}
\end{table}

\subsection*{Lack of Standardized Benchmarks}
Moreover, the lack of publicly available datasets leads to the dearth of standardized benchmarks for evaluating speech-processing tasks on children's audio. 
Most of the evaluation of signal processing tasks has been conducted using in-house datasets owned by researchers. Some small corpora are publicly accessible, often featured in different ML challenges and competitions, as shown in Table~\ref{table:challenges}. 
Without a standard benchmark, it becomes difficult to fairly compare different algorithms or establish a baseline for improvements. 
Small corpora may not fully capture the range of complex real-world acoustic conditions, which could impact model performance in diverse settings or broader contexts.
Additionally, limited standardized datasets hinder researchers from performing reproducible studies, which makes it hard for the community to verify model performances and build upon existing work.

\section*{Review of Algorithms and Recent Advancements in Speech-Processing Tasks}

Early speech processing heavily relied on statistical models, such as the Gaussian Mixture Model (GMM) and Hidden Markov Model (HMM), due to the constraints of smaller labeled datasets and computation resources. With the advent of deep learning (DL), researchers transitioned to DL-based models in a supervised learning setting by leveraging abundant labeled data, e.g., the 960-hour LibriSpeech corpus of adult audiobook recordings for building ASR. These DL models significantly outperformed earlier statistical models. 

\begin{figure}[h]
 	\centering
 	\includegraphics[width=1.0\textwidth]{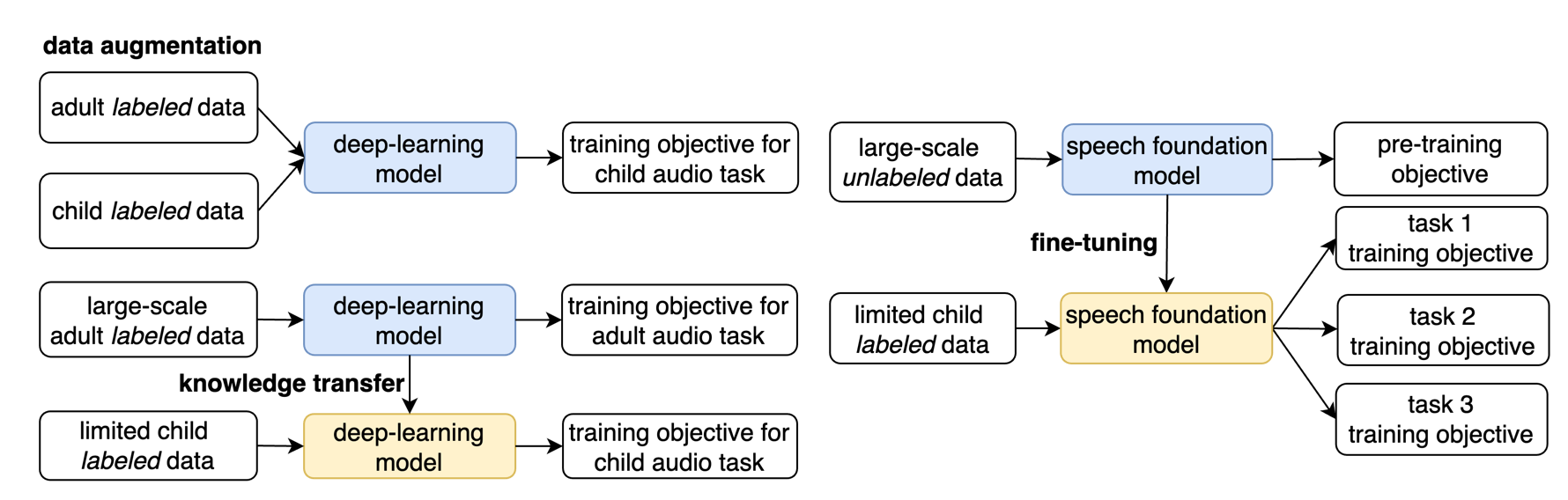}
 	\caption{Major transfer learning techniques used in the automatic analysis of child-centered naturalistic recordings include data augmentation, knowledge transfer through fine-tuning or extracting features from pretrained models initially developed for adult audio to train child audio tasks, and fine-tuning speech foundation models to handle a range of downstream speech processing tasks.}
 	\label{fig:transfer_learning}
 \end{figure}

To address the challenges of analyzing child-centered naturalistic recordings, transfer learning is a key approach to alleviating the data sparsity problem in labeled children's audio data.
 Figure~\ref{fig:transfer_learning} presents a summary of major transfer learning techniques used in past studies for automatically analyzing child-centered naturalistic recordings. 
 Transfer learning in supervised learning leverages knowledge from a source task $\mathcal{T}_s$ trained from domain $\mathcal{D}_s$ to improve performance on a target task $\mathcal{T}_\tau$ from domain $\mathcal{D}_\tau$. 
 A domain $\mathcal{D}$ is defined as  $\mathcal{D} = \{X, P(X)\}$, where $X$ is the feature space that defines input data points and $P(X)$ represents the marginal probability distribution over feature space $X$. In speech processing related studies, $X$ denotes either speech input or embeddings extracted from pre-trained models, $P(X)$ describes how data points are distributed in the feature space, where $X$ may represent different feature types depending on the processing stage.
 In this article, source domain often refers to the domain trained on adult datasets $\mathcal{D_\text{adult}}$, and target domain often refers to domain intended for improvement on child datasets $\mathcal{D_\text{child}}$. 
Simple transfer learning techniques settings include combining child audio data with adult audio data to train models for analyzing child audio. Let $\mathcal{D_\text{adult}} = \{(x_i^{\text{adult}},y_i^{\text{adult}})\}_{i=1}^{N_\text{adult}}$ and $\mathcal{D_\text{child}} = \{(x_i^{\text{child}},y_i^{\text{child}})\}_{i=1}^{N_\text{child}}$ represent adult and child audio datasets respectively, where $N$ represents number of samples and $y_i$ is the label corresponding to $x_i$ (e.g., speaker label for classification). The training loss for a ML model is
\[\mathcal{L} = \sum_{i=1}^{N_\text{adult}}  \ell (f(x_i^{\text{adult}}; \theta),y_i^{\text{adult}}) + \sum_{i=1}^{N_\text{child}}  \ell (f(x_i^{\text{child}}; \theta),y_i^{\text{child}}),\]
where $\ell$ is the loss function (e.g. cross-entropy loss), $f$ is the model with learnable parameters $\theta$.
Advanced transfer learning techniques leverage knowledge transfer from pre-trained models developed using adult audio.
Fine-tuning pre-trained models on adult datasaets using child audio datasets is a common approach, and the model parameters are updated via gradient descent as follows:
\[\theta \leftarrow \theta_{\text{adult}} - \eta \nabla_{\theta} \mathcal{L}_{\text{child}}, \]
where $\theta$ is the updated model parameters, $\theta_{\text{adult}} $ denotes the parameters of the pre-trained model, $\eta$ is the learning rate, and $\mathcal{L}_{\text{child}}$ denotes the loss function for training child audio task.
Besides fine-tuning, features extracted from the pre-trained model using adult datasets can be used to train child audio task. Let $\phi(x_{\text{child}}; \theta_{\text{adult}})$ denote the feature extractor for the pre-trained model, where $x_{\text{child}}$ is child audio input and $\theta_{\text{adult}}$ is the model parameters for the pre-trained model. The extracted features $z_i^{\text{child}} = \phi(x_i^{\text{child}}; \theta_{\text{adult}})$ are used to train a classifier $g$ for analyzing child audio. 
The classifier's loss function is:

\[
\mathcal{L}_{\text{child}} = \sum_{i=1}^{N_{\text{child}}} \ell(g(z_i^{\text{child}}), y_i^{\text{child}}).
\]

To alleviate the requirement of abundant labeled data for supervised learning, speech foundation models (SFMs) were recently invented via (1) \textit{self-supervised learning} by pre-training on up to tens of thousands of unlabeled audio, such as wav2vec 2.0, HuBERT, and WavLM, or (2) \textit{weakly supervised learning} with a large amount of weakly labeled audio, such as Whisper. 
Among these self-supervised learning models, wav2vec 2.0 learns audio representations by first converting raw waveforms into latent feature embeddings. These representations are then passed to both a quantizer and a transformer network. The quantizer maps each latent embedding to a specific learned speech unit from a set of predefined quantized units. Meanwhile, a portion of the latent representations is masked before being input into the transformer network. During training, a contrastive loss is applied to the transformer’s outputs, guiding the model to predict the target audio segment based on the surrounding audio context.
For weakly supervised learning model, Whisper's pre-training process involves learning from a large-scale multilingual and multitask dataset of 680,000 hours of audio-text pairs using an encoder-decoder based sequence-to-sequence framework. The audio input is processed by an encoder to extract latent representations, which are then passed to a decoder to generate the corresponding transcription or text output. The weakly labeled data means the text labels may not be perfectly aligned with the audio, which requires the model to learn contextual relationships between audio and text. 
This approach enables Whisper to handle diverse speech processing tasks such as speech activity detection, multilingual ASR, and speech translation while being robust to noisy or imperfectly labeled data.

Pre-trained SFMs are trained on large and diverse audio datasets and can be fine-tuned with a limited amount of labeled data to handle a wide range of downstream speech-processing tasks.
During supervised fine-tuning, the model may be optimized for a set of downstream tasks $\mathcal{T}$.
For each task $\tau \in \mathcal{T}$, a limited amount of labeled data is provided, represented as $\mathcal{D}_\tau = \{(x^\tau_i, y^\tau_i)\}_{i=1}^{N_\tau}$. The pre-trained model extracts features $h = \phi(x; \theta_{\text{SFM}})$, where $x$ is the audio input and $\theta_{\text{SFM}}$ are parameters learned during pre-training. A task-specific layer $g(h; \psi_\tau)$, such as a simple feed-forward network, convolutional neural network (CNN), or recurrent neural network (RNN), uses these features to predict $\hat{y}= g(h; \psi_\tau)$.
Fine-tuning optimizes the task-specific loss $\mathcal{L}_\tau$, updating both $\theta_{\text{SFM}}$ and $\psi_\tau$ via gradient descent. For multi-task fine-tuning, the parameters are updated separately for each task $\tau \in \mathcal{T}$:
\[
(\theta_{\text{SFM}}, \psi_\tau) \leftarrow (\theta_{\text{SFM}}, \psi_\tau) - \eta_t \nabla_{(\theta_{\text{SFM}},\psi_\tau)} \mathcal{L}_\tau,
\]
where $\eta_t$ is the learning rate for task $t$. 
SFMs serve as a unified framework for multi-task learning and offer benefits beyond supervised learning, which requires training individual models for each specific task. These benefits of SFMs make them a great solution for building child speech processing applications. 

In the following sections, we define several key speech processing tasks and applications for analyzing long-form child-centered naturalistic recordings, including \textit{speaker diarization/speaker type classification, vocalization classification}, and \textit{linguistic unit count and speech recognition}. 
We will also review two emerging tasks that have started to gain attention in recent years and show significant potential for future applications in naturalistic long-form recordings to enhance our understanding of child development, including \textit{speaker verification of infant cry} and \textit{language identification and diarization in CDS}. Table~\ref{table:challenges} presents descriptions of important datasets that are associated with relevant tasks and ML challenges. For each task, we first present an overview of essential algorithm pipelines. Due to data-sharing constraints, a standard benchmark for evaluating these tasks has not been established. We therefore highlight key performance measures reported for notable models developed to tackle these tasks on the specific datasets used in each study. Though different studies are not directly comparable, we aim to provide readers with a sense of state-of-the-art performance in each task.

\subsection*{Speaker Diarization/Speaker Type Classification}
\label{subsection:speakerdiarization}

Speaker diarization, identifying who spoke and when, provides invaluable insights into parent-child interactions by measuring turn-taking dynamics and estimating vocalization counts, which are useful to describe children's experiences and their outcomes.
For diarizing naturalistic family long-form recordings, speaker diarization is often simplified to speaker type classification. Default primary speakers include the key child (the child who wears the audio recorder), adult female, adult male, and other children. Speaker type classification can be considered a computational paralinguistic task. 

Common evaluation metrics for speaker diarization and speaker type classification are different. To evaluate speaker diarization, the diarization error rate (DER) is widely adopted as the benchmark metric.
DER is defined as 
\[\text{DER} = \frac{\text{duration of false alarms + missed detections + speaker confusion errors}}{\text{total duration of reference speech}} \]
where \textit{false alarms} occur when non-speech segments are incorrectly labeled as speech, \textit{missed detections} occur when speech segments are not labeled as any speaker, and \textit{speaker confusion errors} occur when speech segments are assigned to the incorrect speaker. 
Measuring DER can be useful for comparison to the speaker diarization literature, but since the DER metric was originally developed for studies of meetings, interviews, and other conversations between adult speakers in which speech is a majority or near-majority of the audio duration, it overemphasizes timing-related errors and can be disproportionately affected by misclassifications of speech versus non-speech in noisy naturalistic recordings with long silences.
To evaluate speaker type classification, standard classification metrics such as accuracy, precision, recall, and F1 scores are often more appropriate,
because they distinguish between different types of errors that may have very different frequencies in naturalistic recordings. 

\begin{figure}[h]
	\centering
	\includegraphics[width=1.0\textwidth]{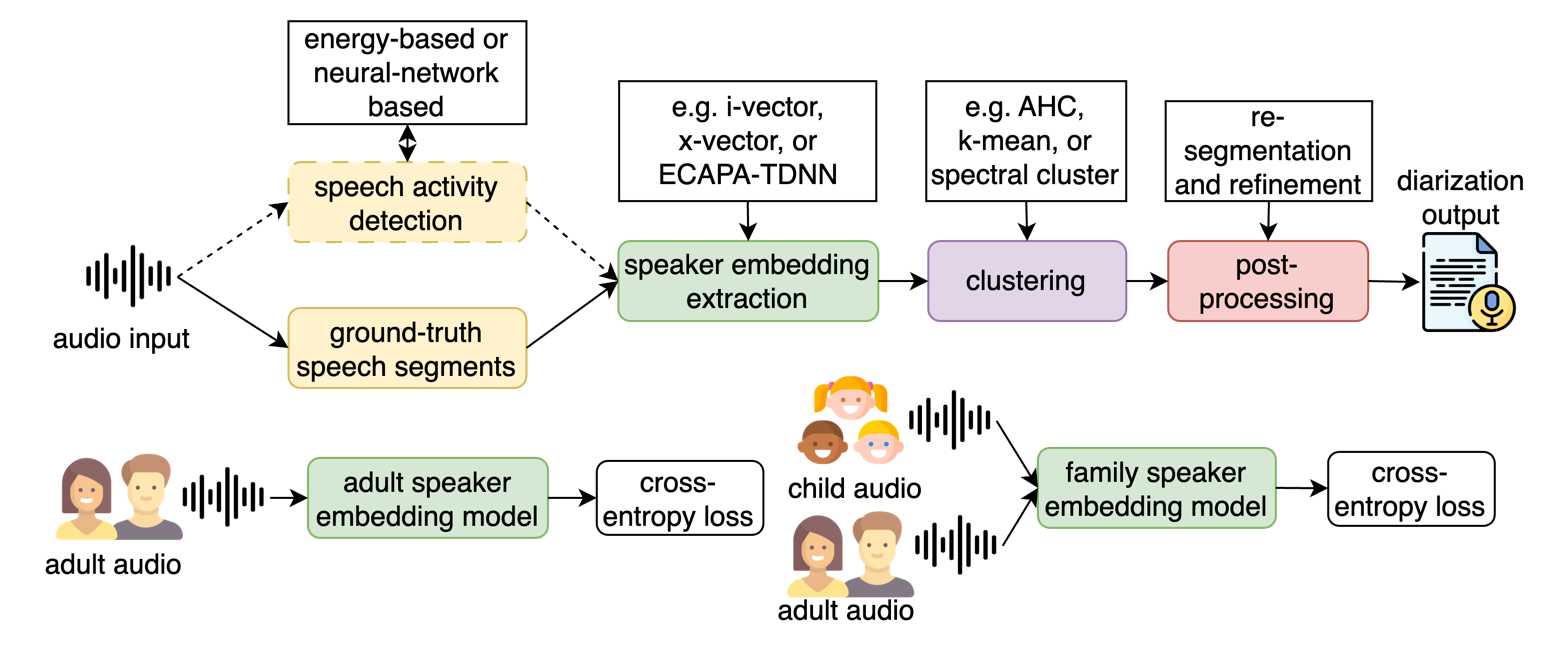}
	\caption{{Overview of clustering-based speaker diarization modules, with common techniques or models specified for each module.} Earlier speech activity detection modules often struggled to capture speech segments in naturalistic recordings, so ground-truth speech segments were used instead. Adult speaker embeddings are pre-trained with adult audio data. Family speaker embeddings are pre-trained with both child and adult audio data.}
	\label{fig:clustering_speaker_diarization}
\end{figure}

Earlier clustering-based speaker diarization algorithms did not directly optimize diarization accuracy and often involved several modules, as shown in Figure~\ref{fig:clustering_speaker_diarization}. First, speech activity detection, which can be an energy-based or neural network-based model, identifies speech segments and removes non-speech segments of silence or background noise. These speech segments are further divided into smaller chunks. Then speaker embeddings are extracted from these smaller chunks to represent speaker characteristics. Speaker embedding models are pre-trained with large-scale speaker datasets using cross-entropy loss to predict speaker identity. For instance, the VoxCeleb dataset contains short speech clips from thousands of speakers and the majority of them are adults.  Earlier speaker embeddings are statistical-based, such as i-vector derived from Joint Factor Analysis or GMMs. Recent DL-based speaker embedding models, including x-vectors and ECAPA-TDNN, showed remarkable performances on the tasks of speaker recognition and verification. The core step, clustering, groups similar speaker embeddings using algorithms such as agglomerative hierarchical clustering (AHC), k-means, or spectral clustering. After clustering, post-processing refinements, such as re-segmentation with HMM, overlap speech detection, and outlier removal of non-speech segments, are applied to enhance the accuracy. The final output consists of time-stamped speaker labels used for evaluation. 
Early speech activity detection models struggled to identify speech segments in highly noisy naturalistic long-form home recordings. Thus, previous studies often used ground-truth speech segments and performed speaker diarization.
Past research also showed adding child speech data, aged 5-10 years old, despite the age mismatch between infants and older children, with adult speech data, is helpful for training robust speaker embeddings for diarizing naturalistic long-form home recordings, as this better captured the distinct acoustic characteristics between speech of adults and children.
These techniques improved clustering-based speaker diarization algorithms on the SEEDLingS corpus used in the DIHARD challenge in terms of DER when using ground-truth speech segments~\cite{krishnamachari2021developing,xie2019multi}. 

 \begin{figure}
	\begin{center}
		\includegraphics[width=0.8\textwidth]{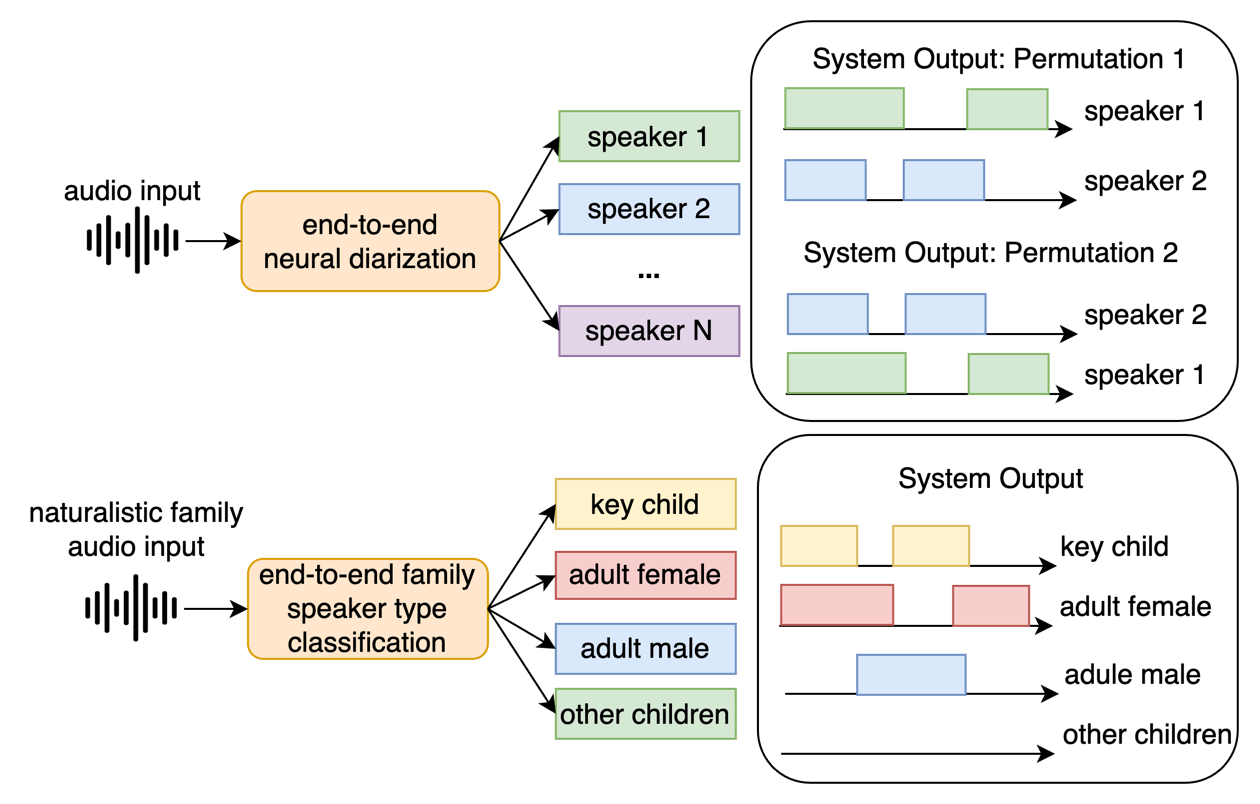}
	\end{center}
	\caption{Overview of algorithm pipelines. Top: End-to-End Neural Diarization (EEND) with two possible permutations of system outputs for two speakers. Bottom: end-to-end family speaker type classification for four primary speakers with one fixed system output. }
	\label{fig:end_to_end_speaker_diarization}
\end{figure}

With the advancement of end-to-end DL model, end-to-end neural diarization (EEND) transformed speaker diarization into a multi-label multi-classification problem~\cite{fujita19_interspeech}. 
EEND introduces a single DL-based model and directly optimizes diarization accuracy by predicting {frame-level} speaker activity{, which allows multiple speaker activities to be active simultaneously and thus enables accurate detection of both single and overlapped speaker detection. EEND processes longer audio contexts (e.g., 1–2 seconds) and predicts binary speaker activity (1 for active, 0 for inactive) for each speaker at a fine temporal resolution (10–20 ms) centered on each frame.}
EEND also introduces a permutation-free objective loss, which minimizes the loss across all possible label assignments, regardless of the predefined order of speakers, to address the ambiguity of arbitrarily assigned speaker labels. 
However, this loss is not applicable to speaker type classification, as speaker types are fixed to specific individuals.
Figure \ref{fig:end_to_end_speaker_diarization} illustrates this difference.
{End-to-end speaker type classification can also handle overlapped speech, similar to how EEND addresses overlapping speakers. Different from end-to-end speaker classification, LENA’s system does not support explicit overlapped speaker detection. Instead, it treats overlapping speech as a separate class and outputs a single additional label, e.g., ``OLN'' in LENA defined label set, without identifying the involved speaker types.}

End-to-end speaker type classification models have been explored in diarizing naturalistic long-form home recordings. Lavechin et al.~\cite{lavechin20_interspeech} built an ML model for open-sourced speaker type classification. Their model was built upon SincNet, a CNN-based architecture that learns low-level speech representations from input waveform, followed by two long short-term memory (LSTM) layers and three feed-forward network layers. The training objective was applying a sigmoid function to {independently} predict speaker activity for five categories of key child, adult female, adult male, other child, and speech from unidentified sources, for multiple audio frames of 16 ms within a larger 2-second context. {This setup allows multiple speaker types to be active at the same time, effectively supporting overlapped speech detection.} This model was trained and tested on the BabyTrain dataset, which consists of 260 hours of naturalistic long-form recordings in 10 languages, sourced from 9 corpora recorded in diverse environments (8 for training and 1 for testing). This model achieved an unweighted average F1 score of 57.3\% across all five categories on the held-out corpora of the BabyTrain dataset, which consistently outperformed the LENA's proprietary technology in identifying each speaker type. 
Besides supervised learning, Li et al.~\cite{li23e_interspeech} pre-trained an SFM model, wav2vec 2.0, on a large-scale dataset of unlabeled 4300-hour naturalistic family long-form recordings, to improve speaker and vocalization type classification task with a limited amount of labeled data. More details are described at the end of \textit{Vocalization Classification} section.

Speaker diarization and speaker type classification algorithms developed for analyzing child-adult interactions in naturalistic long-form recordings are not uniquely tailored to its specific contexts or age groups; instead, these algorithms can be easily extended to diarize child-adult interactions in other contexts. 
For instance, Kothalkar et al.~\cite{Kothalkar2024} developed a lightweight CNN-based DL model for speaker type classification, distinguishing child speech, adult speech, and non-speech segments, to diarize naturalistic recordings of teacher-child interactions in preschool classrooms with children aged 3 to 5 years old. By integrating ASR technology to refine speaker type classification results, they further improved the performance of child-adult speaker diarization. Under a cross-classroom evaluation scheme with 60 hours of data from each of two classrooms, their model achieved an average F1 score of around 80\% for speaker type classification and an average DER of around 40\% for speaker diarization across two classrooms.
Xu et al.~\cite{xu2024exploring} explored fine-tuning several SFMs, such as wav2vec 2.0, Whisper, and WavLMs, for speaker type classification of distinguishing between child and adult speakers in child-parent dyadic interactions with children aged 4 to 8 years, recorded via meeting software, Zoom videos. This study showed that SFMs outperformed clustering-based speaker diarization algorithms in terms of DER score. 

\subsection*{Vocalization Classification}

Vocalization classification plays a crucial role in the automatic analysis of naturalistic recordings, which enable researchers to monitor parent and infant emotional states, assess toddlers' vocal maturity milestones, and understand the influence of parent vocalizations on child language acquisition. {Vocalization classes are defined by child development experts, in order to help them quantify patterns of behavior with theoretical, experimental, or demonstrated clinical importance.  Some of the classification tasks most commonly studied in the literature include infant crying classification, toddler vocal maturity classification, and adult addressee classification. Table~\ref{table:challenges} summarizes those key tasks, their associated labels, and corresponding ML challenges explored in prior studies.}
Vocalization classification requires models to learn nuanced non-verbal cues, acoustic patterns, and paralinguistic features from vocal utterances. Beyond non-verbal cues, this task also requires an understanding of children's phonetic characteristics, rhythmic patterns, and articulatory properties to differentiate between meaningful speech-like sounds (canonical babbling) and non-speech vocalizations (non-canonical babbling). By applying methods commonly used in affective computing, such as speech emotion recognition, models can effectively classify different types of vocalizations into distinct categories.
To tackle this task, utterance-level global features, computed directly from entire utterances or aggregated from frame-level local features through global pooling, are critical for capturing the overall patterns of vocalizations. Local features may complement global features by capturing fine-grained temporal variations and paralinguistic characteristics.

Multiple vocalization classification tasks were featured in Interspeech Computational Paralinguistic (\textsc{ComParE}) Challenges and baseline models often used a feature set composed of utterance-level and frame-level paralinguistic and acoustic features. 
Common baseline feature sets can include thousands of parameters; for example, \textsc{ComParE}\_2016 feature set consists 6,373 parameters. 
In contrast, smaller feature sets, such as GeMAPS (Geneva Minimalistic Acoustic Parameter Set)~\cite{eyben2015geneva}, which contain up to 88 parameters, offer robust performance with a smaller size and greater interpretability for many \textsc{ComParE} tasks. 
GeMAPS is composed of essential features from various acoustic domains such as frequency-related parameters (e.g., pitch and jitter), energy-related parameters (e.g., shimmer and loudness), and spectral-related parameters (e.g., MFCCs).
The feature set extracted per vocalization is then fed into traditional ML-based models, such as Support Vector Machines (SVM), K-Nearest Neighbor, Linear Discriminant Analysis, and Bag-of-Audio-Words, for vocalization classification. 
Replacing traditional ML-based models with DL-based models such as 2-layer feed-word networks, CNN, RNN, and attention-based networks, usually yield better performance. To evaluate vocalization classification, standard classification metrics are used. UAR across all output classes was the primary metric in the \textsc{ComParE} challenges, while F1 score was also commonly reported in previous literature.

\begin{figure}
		\centering
        \includegraphics[width=1.0\textwidth]{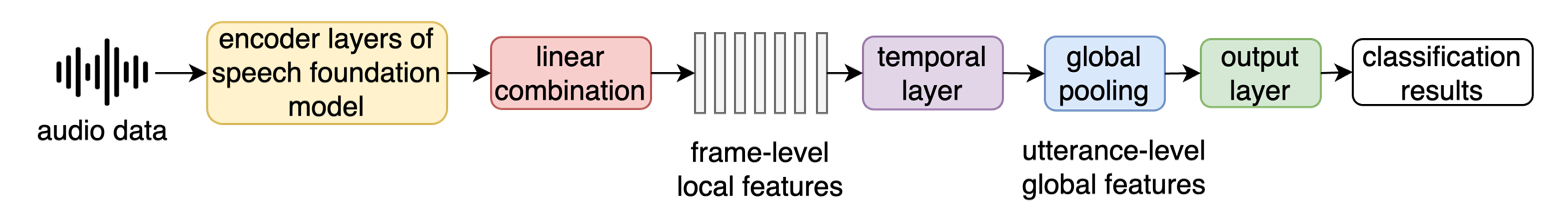}
	\caption{{Overview of a sample vocalization classification algorithm pipeline based on fine-tuning pre-trained SFMs.}
		Layer outputs from the SFM encoders can be linearly combined via weighted sum, average sum, or selected layers. A temporal layer, such as a CNN, RNN, or LSTM, models sequential patterns, followed by global pooling (mean pooling, max pooling, or attentive pooling) to aggregate local features. Global features are fed to an output layer such as a feed-forward network or a linear layer.}
	\label{fig:vocalization_classification}
\end{figure}

Other approaches include using an adversarial autoencoder (AAE) and its variants to learn meaningful latent representations of audio data. AAE is composed of three networks: encoder, decoder, and discriminator. The encoder maps vocalizations to a latent space, while the decoder reconstructs the input, ensuring the latent space captures important features. 
Meanwhile, a discriminator is trained to distinguish between the latent representations generated by the encoder and samples from the prior distribution, such as a Gaussian distribution. The encoder tries to compete with the discriminator by generating indistinguishable latent representations from the prior distribution, which leads to adversarial training.
Once trained, the latent representations can be used for vocalization classification. This approach combines feature learning and classification and effectively learns complex and non-linear vocal patterns, but challenges such as unstable training and mode collapse remain.

Recent advances in fine-tuning SFMs have greatly improved performance in \textsc{ComParE} tasks, which makes this approach a new mainstream. Typically, a linear combination of embeddings from the encoder layers of a pre-trained SFM is used to extract local features, which are then aggregated into global features using global pooling to capture both fine-grained details and overall vocal patterns. An output layer is used to classify vocalizations based on these global features. The SFM encoder layers can be frozen to retain pre-trained knowledge or fine-tuned to adapt to specific tasks, depending on the dataset. Figure~\ref{fig:vocalization_classification} presents an example pipeline for vocalization classification using fine-tuned SFMs.

For infants under 12 months old, crying serves as a vital signal for communicating with caregivers, often indicating infant needs or pathological signals. Numerous studies have explored infant crying through signal processing~\cite{ji2021review}. In this article, we briefly review several key signal processing tasks specifically related to infant vocalization analysis in naturalistic recordings.
Datasets of infant vocalization are typically collected in a variety of settings: laboratories,  homes, and hospitals. 
As a result, DL-based models developed for one of these environments exhibit a performance gap when applied to another due to domain mismatches in acoustic conditions. 
Naturalistic recordings, in particular, are highly noisy, and noise-sensitive DL models trained on clean audio often struggle to perform well when applied to such recordings.
For instance, detecting infant crying {as a binary classification task (crying vs. non-crying)} in long-form naturalistic audio is a foundational task~\cite{9746096}. Yao et al. released a large dataset containing 7.9 hours of annotated crying from naturalistic long-form home recordings collected via LENA. They trained an SVM on {5-second clips from} naturalistic home recordings using acoustic features and deep spectrum features extracted from pre-trained image models, and they achieved an F1 score of 61.3\%.  
They also found DL-models models trained on infant crying collected in the laboratory significantly underperformed when tested on naturalistic recordings, with F1 scores dropping from 65.6\% in laboratory tests to 23.6\% in real-world naturalistic tests, which highlights the challenges posed by real-world environmental noises.
Beyond crying, infant vocalizations provide affective cues that reflect their emotional states. For example, babbling and laughing are associated with neural or positive emotions, while fussing and crying are associated with negative emotions. 
The 2018 \textsc{ComParE} Crying Sub-Challenge classifies infant vocalizations collected in the laboratory into neural/positive babbling, fussing, and crying. Participants explored multiple neural networks trained on various acoustic and paralinguistic feature sets~\cite{huckvale18_interspeech,turan18_interspeech}. 
CNN-based models trained with acoustic features have effectively classified infant vocalizations from open-sourced audio clips and naturalistic recordings~\cite{anders2020automatic,li2021analysis}. Different infant vocalization classes include crying, fussing, babbling, laughing, screeching, and vegetative sounds such as coughing or burping.
Besides CNN-based models, Charola et al.~\cite{charola2023whisper} found that leveraging features from Whisper outperformed CNN or LSTM models in classifying cries from healthy infants as normal, hunger, or pain, and from pathological cases such as deafness or asphyxia. 

From 7 months onward, infants start producing, in addition to cries and laughter, two subtypes of speech-like vocalizations: canonical (e.g., ``ba") and non-canonical (with non-adult-like transitions between vowels and consonants, e.g. ``eeeee"). Such early vocal productions are important markers for the early identification of potential speech and language disorders. 
For this reason, researchers put together BabbleCor~\cite{cychosz2019babblecor}, a dataset with five classes (crying, laughing, canonical, non-canonical, and ``junk'' - namely errors in the diarization system).  
In the 2019 \textsc{ComParE} Challenge, the best-performing system mitigated a strong imbalance across the 5 classes by incorporating synthetic laughter data generated by an AAE network. By training an SVM with multiple paralinguistic and acoustic feature sets, the system achieved a UAR of 62.4\% on the test set~\cite{yeh19_interspeech}. 
Until recently, Li et al.~\cite{li24j_interspeech} fine-tuned the wav2vec 2.0 model, pre-trained on 4,300 hours of naturalistic recordings of children under 5 years old, achieving competitive performance on the BabbleCor reproducible test subset (91\% of the original 2019 \textsc{ComParE} Challenge data). 
To better distinguish between canonical and non-canonical babbling, they incorporated linguistic features from a separate wav2vec 2.0 phoneme recognition model that was built for toddlers aged 1–4 years, which achieved a 
{phoneme} error rate of 60\% on toddler's speech~\cite{10626416}. Despite this imperfect accuracy, fusing these linguistic features with wav2vec 2.0 features pre-trained on naturalistic recordings achieved state-of-the-art performance with a UAR of 64.6\% on the reproducible BabbleCor test subset.
Recently, the Speech Maturity dataset~\cite{hitczenkospeech} was introduced and extended well beyond BabbleCor in terms of participants and cultural environments. 
Moving forward, this dataset holds the potential to advance the development of more robust models for assessing children's vocal maturity.

Caregivers often communicate with infants and toddlers in CDS, also known as Motherese. CDS is characterized by a higher pitch, exaggerated pitch contours, and simplified vocabulary, and CDS has been shown to be beneficial for promoting the social and cognitive development of children. In contrast, ADS refers to regular speech adults use in everyday communication. A key vocalization task for adults is distinguishing between CDS and ADS. 
In the 2017 \textsc{ComParE} Challenge, Tavarez et al. ~\cite{tavarez17_interspeech} investigated multiple feature fusing methods for paralinguistic and acoustic feature sets and outperformed \textsc{ComParE} baselines. 
These methods included early fusion, where multiple feature sets are concatenated into a single vector, and late fusion, which combines output scores from multiple classifiers such as SVM and GMM.
Additionally, CNN-based end-to-end models, pre-trained on ImageNet that contains millions of images, have been applied to mel-spectrogram-based features for classifying to classify CDS versus ADS across multiple languages, including English, Spanish, Arabic, and Chinese sourced from multiple open-sourced or private datasets~\cite{aljarb2023machine}. 
Futaisi et al.~\cite{10096728} combined multiple CDS and ADS datasets from naturalistic long-form recordings across different cultural environments and laboratory settings, with each corpus annotated according to its own protocol. They explored multi-task learning and domain adversarial training to mitigate differences across corpora. However, the study observed limitations in model performance when addressing domain mismatch. This highlights the need for a consistently annotated dataset to determine whether a single model can generalize across diverse languages, cultures, and data collection environments. In the future, incorporating linguistic features extracted from multilingual SFMs may be beneficial for this task, as language style and vocabulary usage differ significantly between CDS and ADS.
Beyond classifying CDS vs. ADS, other parent vocalization types, such as laughter and singing, reflect their positive emotional states that could also play a significant role in children's emotional development. Relevant weakly labeled adult vocalizations can be found in open-sourced audio datasets. 
However, limited studies have focused on classifying adult vocalizations in naturalistic recordings, and a few studies applied CNN-based and attention-based networks for this task~\cite{li2021analysis}. 

To unify computational paralinguistic tasks such as speaker type and vocalization classifications in the automatic analysis of long-form naturalistic recordings, SFM offers a unified framework capable of handling multiple downstream tasks efficiently, even with limited labeled data. Li et al.~\cite{li23e_interspeech} pre-trained an SFM model, wav2vec 2.0, on a large-scale dataset of unlabeled 4300-hour naturalistic family long-form recordings, followed by fine-tuning on a smaller 36-hour in-house labeled dataset. 
The model utilized weighted sum wav2vec 2.0 features and three simple 2-layer feedforward output layers to classify 1-second audio frames within a 2-second context for multitask learning. It achieved robust performance, with an F1 score of 78.2\% for speaker type classification and 70.1\% and 68.8\% for parent and infant vocalization classification, respectively, on a 2-hour in-house test dataset using a leave-one-family-out evaluation scheme.
Speaker type classification includes four primary speakers {without overlapped speaker detection}: key child, adult female, adult male, and other children. Infant vocalizations were classified into crying, fussing, and babbling. Parent vocalization was classified as CDS, ADS, laughter, and singing.
Leave-one-family-out scheme ensures the training and testing datasets don't have overlapped families. This study demonstrated that pre-training wav2vec 2.0 on naturalistic recordings is superior to pre-training on adult speech datasets, such as Librispeech corpus, for family audio analysis tasks, including speaker type classification and parent and infant vocalization classifications. 
Although the system demonstrated robust performance on a smaller testing dataset, further validation is necessary to justify its effectiveness across diverse naturalistic contexts.

\subsection*{Linguistic Unit Count and Speech Recognition} 

Language development researchers investigate not only \textit{who talks to the child} but also \textit{how much language children receive}. This interest in input quantity stems from evidence linking the amount of speech children hear to their later language skills,
In this section, we first review algorithms for counting the number of linguistic units addressed to the child, typically in words. Other types of linguistic units may include phonemes and syllables. Second, we examine how researchers have started leveraging ASR systems to analyze the quantity and content of children's linguistic environments, including the child's own production.

One key measure of linguistic input is the adult word count (AWC), which represents the number of words spoken by adults near the child during a specified timeframe, typically over an entire day at home or daycare.
From a technical viewpoint, estimating AWC, or count of any other linguistic units, presents significant challenges because spoken words flow together without clear acoustic markers separating them. Until recently, researchers primarily relied on LENA's proprietary software to estimate AWC. LENA's AWC estimate involves two main steps \cite{xu2008signal}. First, it segments the audio into broad speaker categories to identify adult speech segments. Then, for each adult segment, it estimates the word count using a least-squares linear regression model, which considers three input variables: the segment length and the numbers of consonants and vowels (with the latter two detected using the English Sphinx 
{phoneme} recognizer). 
A more recent alternative, Automatic LInguistic unit Count Estimator (ALICE), uses a similar 2-step approach but offers several advantages \cite{rasanen2021alice}. ALICE expands beyond word counting to estimate the number of phones and syllables produced by adults, and importantly, it has been optimized for multiple languages including Tseltal, Yélî Dnye, Argentinian Spanish,  American, Canadian, and British English. This multilingual capability is achieved through a pre-trained multilingual phone recognizer 
and end-to-end CNN-based syllable count estimator \cite{seshadri2019sylnet} to extract various features. These features are then used as input variables in the least-squares linear regression model, whose parameters are optimized on a multilingual training set. 

Systems to estimate the number of linguistic units addressed to the child are typically evaluated using the absolute relative error (ARE) between automatic and human estimates, which is defined as: 
\[ ARE(s) = \dfrac{\lvert N_{\text{hypo,s}} - N_{\text{ref,s}} \rvert}{N_{\text{ref,s}}} \]
where $s$ is an audio clip, $N_{\text{hypo},s}$ is the number of linguistic units estimated for this clip, and $N_{\text{ref,s}}$ is the actual number of linguistic units. Instead of ARE, which measures whether the absolute unit counts are accurate, correlation scores can be used to measure whether the system can correctly identify which subjects have higher/lower counts compared to others. Using a leave-one-corpus-out procedure to evaluate the system's generalization performance to unseen data, ALICE demonstrated better word count estimation compared to LENA across four English-speaking child-centered recording corpora, achieving an absolute 17\% lower average error rate \cite{rasanen2021alice}. Note that both systems were evaluated using real-world conditions: ALICE used the speaker type classification algorithm \cite{lavechin20_interspeech} for adult speech detection, while LENA used its proprietary diarization system. This realistic evaluation approach better reflects performance in practical applications where oracle diarization is not available. 

Instead of simply counting linguistic units, a system identifying the content of these units would be even more useful. Moreover, automatically transcribing young children's speech into words (or phonemes for non-lexical vocalizations) would allow researchers to precisely track language development. Transcribing caregivers' speech would reveal the linguistic input from which children learn. 
While research on ASR for children is growing, most existing corpora focus on children between 5 and 18 years of age recorded in relatively clean conditions \cite{bhardwaj2022automatic}. 
Despite the potential value of ASR systems for early language acquisition research, surprisingly little work has focused on developing robust solutions for this specific context. We look forward to future thorough evaluations of applying the latest advanced ASR model to this task.

In the absence of systems specialized for processing child-centered long-form recordings, preliminary efforts have focused on building ASR systems to transcribe teacher-child interactions in preschool classrooms. ASR systems are typically evaluated at word error rate (WER), defined as:
\[ WER = \dfrac{S + D + I}{N}\]
where S is the number of substitutions, D is the number of deletions, I is the number of insertions needed to transform the system's output into the reference transcription, and N is the total number of words in the reference.
Lileikyte et al.~\cite{lileikyte2022assessing} applied HMM, DNN, and LSTM models to perform ASR on a 15-hour dataset of young children aged 3–5 years in naturalistic preschool classrooms, and they achieved the best WER of 57.8\%.  
Recent studies have used Whisper to enhance the ASR performance.
When applied directly to full recordings, Whisper tends to hallucinate speech in segments containing no vocal activity. 
One emerging pipeline overcomes this issue by first applying a speaker type classification system to identify speaker activity from both adults and children, then using Whisper to transcribe these speech segments. 
In \cite{Sun2024WhoSW}, Sun et al. applied this pipeline to naturalistic recordings collected in US preschool classrooms using teacher- and child-worn microphones. Performance is reported on 2 hours of audio annotated by a human expert, including 85 minutes from child-worn microphones with 4 children between 3 and 5 years old and 25 minutes from teacher-worn microphones with 2 teachers. The authors found an average WER of 14.7\% for adult speech and 15.0\% for child speech, which indicates that 85\% of the words were identical between the human and the automatic transcript on this private corpus. Long et al.~\cite{Long2024TheBD} reported comparable performance on the BabyView-Preschool portion dataset. On this corpus, the pipeline achieves a WER of 12\% for adult speech based on 1,298 utterances, and 18\% for child speech, including 877 utterances from children aged 3 to 5 years.
Results from these two independent studies suggest that automatic transcripts show a high level of agreement with human transcripts in preschool classroom settings.

The performance is somewhat lower in naturalistic home environments. On the BabyView-Home dataset, also collected from head-mounted cameras but this time in American English-speaking homes, results indicate a WER between 30 and 37\% on adult speech and 21\% and 111\% for child speech (with higher WER for younger children) \cite{Long2024TheBD}. Beyond variability in child-directed and child-produced speech, far-field speech and background noise remain a significant challenge for ASR systems: Testing on 2 hours of annotated audio from SEEDLingS~\cite{seedlings}, Lavechin et al.~\cite{lavechin2023babyslm} reported an average WER of 17\% in clean recording conditions (signal-to-noise ratio in [12, 24] dB range), while the average WER escalated to 62\% in noisy recording conditions (signal-to-noise ratio in the [-9, -4] dB range). 
Li et al.~\cite{10626416} reported a similar level of error rates for recognizing toddlers' vocalizations in noisy naturalistic recordings.  A phoneme recognition model was built for recognizing toddlers' vocalizations by fine-tuning SFMs including wav2vec 2.0 and HuBERT. The model used three-level fine-tuning on speech from adults, older children aged 8-10 years, and toddlers aged 1–4 years to gradually mitigate the age mismatch. It achieved a 
{phoneme} error rate of 60\% on 25 hours of toddler speech from two children in a longitudinal home study within the Providence corpus~\cite{demuth2006word}. 

In summary, these recent studies underscore both the potential and current limitations of automated speech processing for child language research. While structured environments like preschool classrooms show performance approaching human-level accuracy, significant challenges remain for naturalistic home environments and noisy conditions.

\subsection*{Speaker Verification of Infant Cry}

Speaker verification determines if two vocalization clips belong to the same speaker, which is often useful for security and authentication. Recently, speaker verification has been extended to verifying infant speakers based on their crying sounds. Infant cries are known to reflect their needs, and they also contain signs related to health issues, especially neurological and respiratory issues. Clinicians have observed notable differences in crying patterns between healthy babies and those with neurological injuries~\cite{kheddache2015resonance}. Speaker verification of infant crying is beneficial for the early detection and monitoring of neurological injuries in newborns. 
Additionally, from a technological perspective, the algorithms developed for this task can help verify the identity of the infant who wears the recorder and perform target speaker analysis in long-form recordings. Target infant speaker analysis focuses solely on analyzing the vocalizations of target infants while ignoring vocalizations from other speakers. This is useful for distinguishing the target infant from other children with similar acoustic characteristics, such as twins or young siblings.

\begin{figure}
	\centering
	\begin{subfigure}[b]{1.0\textwidth}
		\includegraphics[width=1\linewidth]{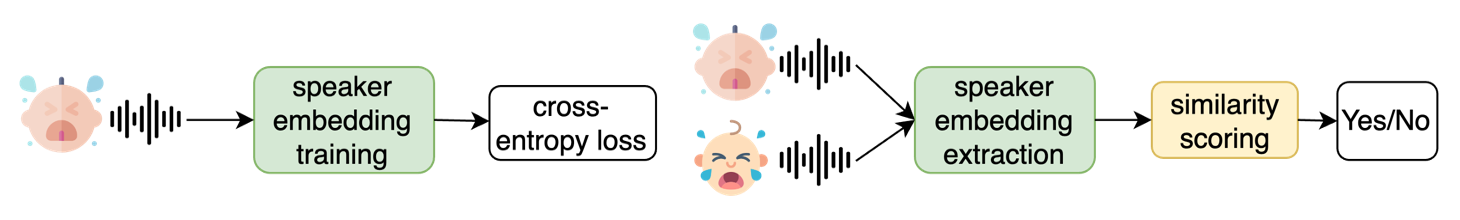}
		\caption{Speaker embedding-based model}
		\label{fig:speaker_verification_embed} 
	\end{subfigure}
	
	\begin{subfigure}[b]{1.0\textwidth}
		\includegraphics[width=1\linewidth]{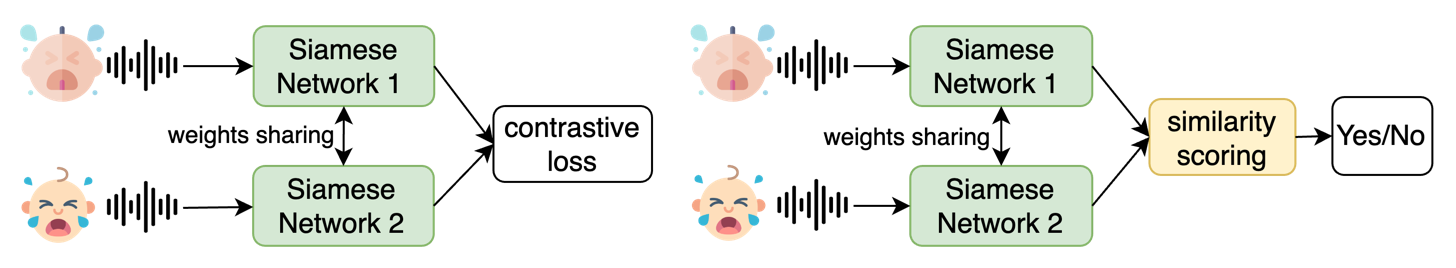}
		\caption{Siamese networks-based model}
		\label{fig:speaker_verification_siamese}
	\end{subfigure}
	\caption{Two common algorithms for speaker verification: the training stage (left) and the verification stage (right).}
	\label{fig:speaker_verification}
\end{figure}

Figure~\ref{fig:speaker_verification} shows two common algorithms for speaker verification including speaker embedding-based model and Siamese networks-based model. 
For speaker embedding-based model shown in Figure~\ref{fig:speaker_verification_embed}, a speaker embedding model, such as x-vector or ECAPA-TDNN, is first pre-trained on large-scale speaker datasets using cross-entropy loss to predict speaker identities. At the verification stage, speaker embeddings are extracted from an input pair of vocalization clips, and a similarity score, such as cosine similarity, is computed between the pair of embeddings. A threshold is then selected to determine if the clips belong to the same speaker. 
For the Siamese networks-based model shown in Figure~\ref{fig:speaker_verification_siamese}, two sub-identical networks, sharing the same architecture and weights, are trained to compute speaker embeddings for input pairs of vocalization clips. The training objective uses contrastive loss, which maps embeddings from the same speaker closer together and those from different speakers further apart. In the verification stage, similarity scores are calculated to determine whether a pair of clips belongs to the same speaker. 
The standard evaluation metric for speaker verification is the equal error rate (EER), which is the point on the receiver operating characteristic (ROC) curve where the false acceptance rate equals the false rejection rate. EER is the threshold at the system is equally likely to incorrectly accept a non-matching pair as it is to wrongly reject a matching pair. The lower the EER, the better the system performance.

CryCeleb, a large dataset consisting of 26k infant crying clips collected over 6 hours from 786 newborns in multiple hospitals by Ubenwa, was recently introduced in 2023 for the infant speaker verification challenge~\cite{budaghyan2024cryceleb}. The top-performing model was based on fine-tuning ECAPA-TDNN models that were pre-trained on the VoxCeleb speaker dataset on the CryCeleb dataset. The model achieved a 25.8\% EER on the test dataset of 25k pairs of audio clips from 160 infants. Notably, ECAPA-TDNN achieved 0.87\% EER on the VoxCeleb dataset,
which suggests a significant performance gap between adult and infant speaker verification. 
Siamese-based models have also been explored but didn't achieve desirable performances, likely due to the smaller dataset size of CryCeleb. Further research in this emerging area are needed to improve model performance.

\subsection*{Language Identification and Diarization}
In multilingual families, adults often engage in code-switching during everyday communication with children. Code-switching, the practice of alternating between two or more languages within a conversation, is common in these settings and reflects the dynamic interplay of languages in naturalistic communication. Automatically analyzing these language dynamics requires advanced multilingual speech processing models to handle two critical tasks, including language identification and language diarization.
Language identification recognizes the language spoken in a monolingual speech clip. In contrast, language diarization extends this task to multilingual speech clips by automatically segmenting multilingual speech when language switches occur and identifying languages for each segment. 
Language identification has been extensively studied in adult speech processing with development of robust algorithms and applications. For example, it's commonly used as an auxiliary task to improve multilingual ASR by identifying the language present in a speech segment. This foundational task provides language-specific information to multilingual speech processing systems. 
In contrast, language diarization remains a relatively less-studied area, even within adult speech processing algorithms. Challenges such as the lack of high-quality labeled datasets and the lower practical demand for transcribing multilingual conversations in real world have limited its development. However, language diarization has gained increasing attention in recent years due to its importance in studies involving code-switching in bilingual children and understanding children’s linguistic environments for developmental research.
Techniques developed for speaker diarization, such as clustering-based and end-to-end pipelines, can be adapted to language diarization. These methods segment and classify speech based on language rather than speaker identity. For instance, clustering-based methods group speech segments with similar acoustic features, while end-to-end systems directly predict language transitions.
Both language identification and diarization are essential for understanding the interplay of languages and cultures in parent-child interactions. Analyzing these dynamics provides valuable insights into how children are exposed to and acquire multiple languages in naturalistic settings.

To evaluate language identification, standard classification metrics such as accuracy and F1 scores are widely applicable.
Specifically for bilingual code-switching datasets~\cite{chua23_interspeech}, EER and balanced accuracy (BAcc) are frequently employed. BAcc is particularly valuable for handling imbalanced class distributions. Unlike standard accuracy, which treats all samples equally, BAcc ensures that both majority and minority classes are given equal weight. It is calculated as the average recall for each class:
\[\text{Balanced Accuracy} = \frac{1}{2} (\frac{TP}{TP+FN}+\frac{TN}{TN+FP})\]
where TP is true positive, TN is true negative, FN is false negative, and FP is false positive.
For language diarization, where recordings may contain multiple language labels, evaluation often uses Language Diarization Error Rate (LDER), an adaptation of the DER introduced in the context of speaker diarization The LDER is defined as:
\[\text{LDER} = \frac{\text{false alarms + miss detections + language errors}}{\text{total referenced time}} \]
where false alarms refer to the duration where nonspeech segments are mistakenly labeled as any target language, missed detections represent the duration where active speech regions are incorrectly labeled as nonspeech, and language errors correspond to the duration where an incorrect language tag is assigned to an active speech region~\cite{chua23_interspeech}. In cases where language diarization is treated as a classification task, slice-wise labels can be assigned through majority voting, allowing classification metrics to be used as an additional criteria~\cite{wang2024bridgingChildLDLIDMethod}.

 Recently, the MERLIon CCS challenge introduced a new English-Mandarin code-switching dataset featuring over 30 hours of manually transcribed CDS in parent-child book-reading recordings at diverse home environments. The challenge proposed both tasks of language identification and diarization~\cite{chua23_interspeech}. 
 To tackle the challenge, leveraging multilingual SFMs or pre-trained multilingual language identification systems, such as Whisper or XLSR-53--a wav2vec model pre-trained on 56k hours of 53 languages--has shown to be beneficial~\cite{praveen23_interspeech, wang2024bridgingChildLDLIDMethod, gupta23_interspeech}. 
 Phonological-related features are highly discriminative among different languages, and previous studies explored incorporating such features to enhance language diarization performance.
 Shahin et al.~\cite{shahin23_interspeech} used wav2vec-based model to first learn language-dependent phonological features by fine-tuning first on detecting manners and places of articulation followed by classifying between English and Mandarin speech segments to tackle the task of language identification.
 Wang et al.~\cite{wang2024bridgingChildLDLIDMethod} explored leveraging a combination of phonetic and phonotactic embeddings, extracted from the XLSR-53 model, as speech representations at varying resolution levels. Phonetic embeddings are considered lower-level phonological representations that capture individual phoneme characteristics, while phonotactic embeddings, representing higher-level phonological features, are extracted from higher transformer layers and capture the sequential and combinatorial rules of phonemes
When pre-trained on an out-of-domain adult code-switching and fine-tuned with MERLIon CCS dataset, the model achieved a BAcc of 53.19\% for language identification and a LDER of 32.99\% for language diarization on the MERLIon test set. 

Beyond the MERLIon CCS challenge, recent research started exploring the task of language identification task in naturalistic home recordings. Dutta et al.~\cite{dutta24_odyssey} analyzed the CHILDES Spanish-English Hoff Corpus, which contains naturalistic adult-child bilingual interactions collected from children aged 2-3 years old. The proposed model integrated both language identification (Spanish vs. English) and speaker type classification (Adult vs. Child) into a unified multi-task learning framework to process audio samples of 1-3 seconds. By pre-training XLSR-53 with adult Spanish and English speech and fine-tuning on Hoff corpus, the model achieved a BAcc of 78.92\% (Adult 83.9\%; Child 67.64\%) and an EER of 19.57\% (Adult 15.28\%; Child 29.59\%) on language identification task. In the future, algorithms developed for language identification and diarization could be extended to naturalistic recordings, discovering valuable insights into the linguistic environments of multilingual children.

\section*{Opportunities and Future Directions}
\subsection*{Establish Standardized Benchmarks}
{Previous research has made initial efforts to investigate important tasks in analyzing long-form naturalistic recordings, but current performance still lags behind the state-of-the-art achieved by speech-processing models developed for adults. Task-specific evaluations have included speaker diarization and vocalization classification for infants, toddlers, and caregivers, as outlined in Table~\ref{table:challenges}. Typically, previous work has proposed task-specific benchmark datasets, which often covered a wide range of home environments, languages, and demographic backgrounds}.
{However, an ideal} benchmark dataset should include consistent annotations across multiple tasks, such as speaker types, vocalization types, and transcripts of parent and child speech.  Similar to how the LibriSpeech dataset serves as a benchmark for speech recognition, such a dataset would enable researchers to systematically evaluate and compare the performance of their algorithms. 
{Such a benchmark would support the comparison of both task-specific models and unified frameworks capable of handling multiple tasks, supporting reproducible models and facilitating the measurement of progress in the field.}


\subsection*{Expand Multilingual Studies} 
Most published studies focus on children in English-speaking countries. To gain a deeper understanding of parent-child interactions and child development across various languages, cultures, and countries, there is a need for increased data collection and analysis in diverse linguistic contexts. 
{Cultural norms and practices significantly shape parenting approaches and influence CDS. Additionally, language acquisition affects children's vocal development. For example, children gradually acquire certain sounds of phonological patterns in their native languages. Therefore,} expanding multilingual studies 
questions worth exploring include: {for distinguishing CDS vs. ADS,} how well do ML models trained on data from English-speaking countries 
in non-English speaking countries? Do multilingual SFMs outperform monolingual SFMs in distinguishing multilingual CDS from ADS? What features do these models rely on to differentiate CDS across culturally and linguistically diverse settings?  {For toddler phoneme recognition and vocalization classification, how sensitive is the model to language-specific phonological structures?}

However, conducting multilingual studies may present {several} challenges. {Linguistic differences, such as tonal variation, morphology, and phoneme inventory, can affect the generalization of existing SFMs. Moreover, regulations regarding the ethical treatment of human subjects may vary across countries and require localized research protocols. Annotating caregiver-child interactions and vocalizations may also require culturally sensitive coding schemes to capture behaviors specific to a given cultural context.} These challenges, {however,} also provide an opportunity for collaboration among researchers
and policymakers to develop practical methodologies for analyzing parent-child interactions in diverse cultural and linguistic contexts. By addressing these issues, multilingual studies can contribute to a deeper, more equitable understanding of child development across the globe.

\subsection*{Build Robust Speech Enhancement Models} 
Enhancing the analysis of naturalistic recordings requires robust speech enhancement techniques for single-channel audio. Similar to the cocktail party problem, where humans can distinguish between different sound sources and background noise, improving speech enhancement techniques will allow machines to better isolate individual audio sources in complex environments with overlapping sounds. 

A key challenge lies in the domain mismatch between major speech enhancement datasets and the characteristics of naturalistic recordings. Most existing algorithms and models are designed and optimized for controlled speech and noise datasets, which do not fully capture the variability, complexity, and noise patterns found in naturalistic long-form home recordings.
One promising approach is to extend existing speech enhancement models by incorporating additional specialized modules tailored to the naturalistic recording domain. These modules can focus on analyzing domain-specific noise patterns, reverberation, or overlapping speech sources. 
Advancing speech enhancement techniques for naturalistic recordings may significantly improve the accuracy and reliability of downstream speech-processing tasks in complex acoustic environments.


\subsection*{Improve Word Count Estimate and ASR Accuracy in Naturalistic Recordings}
Developing robust ASR models that can accurately recognize adult and toddler speech in naturalistic settings would greatly enhance the understanding of adult-child interactions in everyday contexts. State-of-the-art ASR algorithms have achieved high accuracy in transcribing clean adult speech, yet they still struggle with transcribing speech in naturalistic recordings. 
The CHiME (Computational Hearing in Multisource Environments) challenge series has driven advancements in far-field robust ASR, source separation, and speech enhancement technologies.
By leveraging these innovations based on robust front-end speech enhancement and refined back-end ASR engines, far-field ASR models may improve the transcription accuracy of adult speech in naturalistic settings. 

Toddlers' language capabilities are still in the early stages of development. By the age of three, they typically expand their vocabulary to hundreds of words and begin forming three- or four-word sentences. Automatically transcribing or estimating toddlers’ word counts can be highly beneficial for the early detection of speech and language delays. Transcribing toddler speech presents significant challenges due to the appearance of meaningless strings well into the second year of life, and even afterward irregular phonetic pronunciations, incorrect grammar usage, low speech intelligibility, and additional complexities introduced by naturalistic conditions. 
The research community ultimately needs a diverse dataset of transcribed toddler speech to advance this effort, recognizing that obtaining highly accurate transcripts may not always be feasible due to the inherent challenges of low toddler speech intelligibility.
In the future, leveraging noise-robust SFMs with advanced ASR techniques to recognize simple and potentially sparse words produced by toddlers in everyday environments could provide a powerful tool for tracking their language development milestones.



\subsection*{Develop Unified Speech Processing Model and Novel Applications with Spoken Language Models (SLMs)}

SFMs provide an innovative approach to mitigate the need for labeled datasets and build a unified model for tackling multiple tasks of analyzing children's speech.
Meanwhile, large language models (LLMs) have shown remarkable capabilities in understanding multimodal data, such as text, audio, and vision. Emerging research has started applying LLMs to analyze lengthy clinician-child transcripts to support child-oriented tasks, such as predicting activities children are engaged in, assessing their language skills, and identifying clinically relevant traits of autism~\cite{feng2024can}. 

\begin{table}[ht]
\centering
\setlength{\tabcolsep}{2pt}
\resizebox{\textwidth}{!}{%
\begin{tabular}{|l|l|l|l|}
\hline
\textbf{Architecture Type} & \textbf{Input} & \textbf{Output} & \textbf{Use Case Example} \\
\hline
Speech-only LM & Speech & Speech & Children's ASR \\
\hline
Text-only LM & Text & Text & Language assessment based on transcripts \\
\hline
Joint Speech-Text LM & Speech, Text & Speech, Text & Storytelling agents with child-directed feedback \\
\hline
Speech-Aware LM & Speech, Text & Text & Audio understanding of child-centered naturalistic recordings \\
\hline
\end{tabular}%
}
\caption{{Overview of speech and text LM architectures and their sample use cases.}}
\label{tab:slm_usecases}
\end{table}

{Building on the advancements of LLMs in handling multimodal data, spoken language models (SLMs) have emerged as a new paradigm for unifying speech processing tasks within a single system. SLMs integrate speech encoders, such as SFM encoders and audio codec, with the reasoning capabilities of LLMs to perform a broad range of speech-related tasks.
A recent SLM survey by Arora et al.~\cite{arora2025landscape} categorizes various SLM architectures, ranging from pure speech-only language models to joint speech-text models. The multimodal SLMs accept speech and/or text as input and produce speech and/or text as output (speech is included in the input modality, output modality, or both). 
Current SLMs primarily focus on processing adult speech, it should be possible for them to generalize beyond speech to a broader audio domain, including child-centered naturalistic recordings. However, these models may encounter significant domain shifts when applied to children’s speech, largely due to the scarcity of child-centered data in the mainstream speech training datasets used for pretraining. Addressing this challenge typically involves first pre-training a speech encoder and followed by fine-tuning the entire model on in-domain child-centered datasets.
Table~\ref{tab:slm_usecases} presents an overview of text and speech LM architectures, their input and output modalities, and corresponding sample uses.
For example, Speech-only LLMs, or all types of SLMs, may be effective for children’s ASR if their encoders are trained on corpora including children’s speech, while text-only LLMs can support language assessment based on the transcriptions of children's speech. Joint speech and text LM models that generate speech outputs enable more advanced and novel applications, such as building child-directed interactive storytelling agents.
Speech-aware SLMs that generate only text can perform complex audio understanding tasks across diverse contexts when trained on paired speech/audio-text data}. For example, given an audio clip from a naturalistic recording with a sample text prompt, ``\textit{Who is the speaker in the audio? What type of vocalization does the speaker produce? What does the speaker say?}'' The corresponding output text could be: ``\textit{The speaker is an infant. The vocalization type is babbling.}'' 
Beyond tackling paralinguistic computational or speech recognition tasks for adults, 
{SLMs} hold the potential to provide a flexible framework for analyzing complex child-centered audio data, enhancing the understanding of adult-child interactions, and supporting other novel applications.

{Currently, building task-specific models remains essential as large-scale and labeled child-centered naturalistic recordings for training unified SLMs don't exist. The adaptation from a task-specific model to a general-purpose model has underperformed task-specific training in the naturalistic recordings in prior studies~\cite{li23e_interspeech, ryant19_interspeech}. However, these two directions, building specialized or generalized models, need not be mutually exclusive. We encourage researchers to begin by developing specialized models to address immediate research and clinical needs, while also gradually contributing to the long-term goal of building more unified SLMs that can support diverse tasks. This dual focus can help bridge the gap between task specificity and model generalization, offering a practical and forward-looking roadmap for advancing the field.}

\section*{Conclusions}
In this article, we present an overview of the field of automatic analysis of naturalistic recordings of children under three years of age from a signal processing perspective, highlighting its significance and applications. 
We discuss data collection procedures and considerations, including descriptions of common recording devices, annotation procedures, ethical considerations, and widely used data repositories.
We also acknowledge challenges in analyzing naturalistic recordings and review progress in developing speech processing algorithms for five key tasks. A major approach is leveraging transfer learning techniques from models trained on large-scale adult speech or audio datasets to analyze child-centered naturalistic long-form recordings. Despite these advancements, model performance in analyzing real-world naturalistic recordings still lags behind the state-of-the-art achieved by models developed for analyzing adult speech.
We outline five directions for future research, with a critical focus on curating a comprehensive and publicly accessible dataset and establishing benchmarks that can be used to systematically evaluate existing model performance and push technological advancements.
{Of particular note is the exciting direction towards developing a general-purpose and unified SLM that can process a broad range of audio and speech inputs from both adults and children.} 
Furthermore, we encourage ongoing innovation and collaboration between the signal processing community and {researchers in} other relevant fields such as psychology, cognitive science, and linguistics. These {interdisplinary} collaborations can deepen our understanding of early childhood development and enhance the accessibility of speech processing tools. 

%
\section*{Acknowledgment}
NLM and MHJ acknowledge support from the National Institute on Drug Abuse under grant 1R01DA059422.  AC acknowledges the J. S. McDonnell Foundation Understanding Human Cognition Scholar Award, and the European Research Council (ERC) under the European Union’s Horizon 2020 research and innovation programme (ExELang, Grant agreement No. 101001095).

%
\bibliographystyle{IEEEtran}
\bibliography{refs}

%
%
%
%
%
\section*{Biographies}
\label{sec:bio}
%
%

\begin{IEEEbiographynophoto}{Jialu Li}(~\IEEEmembership{Member,~IEEE}) (jialuli@arizona.edu) received her B.S. and Ph.D. degrees in Electrical and Computer Engineering (ECE) from the University of Illinois Urbana-Champaign (UIUC). She was a postdoctoral researcher at UIUC and a visiting scholar at the Languages Technologies Institute, Carnegie Mellow University. She is now an Assistant Professor at the College of Information Science at University of Arizona, Tucson, Arizona, 85743, USA. She is a receipt of the Beckman Graduate Fellowship and Rambus Fellowship at UIUC. Her research focuses on developing AI-powered emerging speech technologies for monitoring infant psychological development and for the early identification of children at risk of autism.
\end{IEEEbiographynophoto}
\vspace{11pt} 

\begin{IEEEbiographynophoto}{Marvin Lavechin} (marvinlavechin@gmail.com) received his Ph.D. degree from École Normale Supérieure, Paris, France. He was a postdoctoral researcher at GIPSA-lab, Université Grenoble Alpes and a visiting researcher at the BabyDevLab of the University of East London. He is now a Simons postdoctoral fellow in the Computational Psycholinguistics lab at the Massachusetts Institute of Technology, Cambridge, MA, 02139, USA. His research focuses on developing AI tools to analyze children's language environments across diverse populations and creates models to simulate language acquisition. Through this interdisciplinary approach, he aims to design machine learning algorithms that better mimic children's learning trajectories and identify mechanisms driving language acquisition.
\end{IEEEbiographynophoto}
\vspace{11pt}

\begin{IEEEbiographynophoto}{Xulin Fan}(~\IEEEmembership{Student Member,~IEEE}) (xulinf2@illinois.edu) received his B.S. and M.S. degrees from the University of Illinois Urbana-Champaign (UIUC), where he is currently pursuing a Ph.D. in Electrical and Computer Engineering. He is a member of the Statistical Speech Technology group at UIUC, Champaign, Illinois, 61820, USA. His research focuses on improving audio representation learning and audio tagging for infant-centered family recordings.
\end{IEEEbiographynophoto}
\vspace{11pt} 

\begin{IEEEbiographynophoto}{Nancy L. McElwain} (mcelwn@illinois.edu) received her B.A. degree from Johns Hopkins University and her Ph.D. from the University of Michigan. She was a postdoctoral fellow at the Center for Developmental Science at the University of North Carolina-Chapel Hill and is now a Research Professor in the Department of Human Development and Family Studies at the University of Illinois Urbana-Champaign, Illinois, 61801, U.S.A, where she directs the Interdisciplinary Lab for Social Development. Her research program focuses on dynamic parent-infant interactions that shape social-emotional development, with the long-term aim to promote positive infant-family relationships and infant mental health through technological innovations.  
\end{IEEEbiographynophoto}
\vspace{11pt} 

\begin{IEEEbiographynophoto}{Alejandrina Cristia} (alecristia@gmail.com) received her B.A. degree from Universidad Nacional de Rosario and her M.A. and Ph.D. degrees from Purdue University. She was a postdoctoral researcher at the Max Planck Institute (Nijmegen) and is now a research director at the Centre National de la Recherche Scientifique (CNRS), at the Laboratoire de Sciences Cognitives et Psycholinguistique (LSCP) cohosted by the Ecole Normale Supérieure, EHESS, and PSL University, Paris 75005, France. Her long-term aim is to shed light on the child language development,  both descriptively and mechanistically, drawing insights from linguistics, psychology, anthropology, economics, and speech technology. She is an ISCA Fellow.
\end{IEEEbiographynophoto}
\vspace{11pt} 

\begin{IEEEbiographynophoto}{Paola Garcia-Perera}
(~\IEEEmembership{Member,~IEEE}) (lgarci27@jhu.edu) received her B.S. degree from Tecnol\'ogico de Monterrey, Campus Estado de M\'exico, her M.S. degree from the University of Manchester, and Ph.D. degree from Universidad de Zaragoza. She was a senior research engineer at Nuance and is now an associate research scientist at Johns Hopkins University, Baltimore, Maryland, 21218, U.S.A. Her interests include speech recognition, speaker recognition, diarization, machine learning and language processing. She has been working on children’s speech; including child speech recognition and diarization in day-long recordings. She is a Member of IEEE.
\end{IEEEbiographynophoto}

\vspace{11pt} 

\begin{IEEEbiographynophoto}{Mark A. Hasegawa-Johnson}(~\IEEEmembership{Fellow,~IEEE}) (jhasegaw@illinois.edu) received his B.S. M.S. and Ph.D. in Electrical Engineering from MIT in 1996.  He was a postdoctoral researcher at University of California, Los Angeles and is now the M.E. Van Valkenburg Professor of Electrical and Computer Engineering at the University of Illinois at Urbana-Champaign, Illinois, 61801, U.S.A. Dr. Hasegawa-Johnson is currently Editor in Chief of the IEEE Transactions on Audio, Speech, and Language Processing. His research interests include automatic speech understanding as an empowerment tool for parents, children, learners, and people with disabilities, and unsupervised learning methods for speech technology. He is a Fellow of the IEEE, a Fellow of the ASA, and a Fellow of ISCA. 

\end{IEEEbiographynophoto}

\vspace{11pt} 


\vfill

\end{document}